\documentclass[11pt]{article}
\usepackage{epstopdf}
\usepackage{color}
\usepackage{subfigure}
\usepackage{amsmath}
\usepackage{amssymb}
\usepackage{graphicx,color}
\usepackage{cite}
\usepackage{enumerate}
\usepackage{amsthm}% theorems and proofs
\newtheorem{proposition}{Proposition}
\newtheorem{definition}{Definition}
\newtheorem{remark}{Remark}
\newtheorem{lemma}{Lemma}
\newtheorem{theorem}{Theorem}

\usepackage{appendix}
\usepackage{amsfonts,mathrsfs}
\usepackage{geometry}
\usepackage{ifthen}
\usepackage{float}
\usepackage{bm}
\usepackage{tabularx}
\usepackage[utf8]{inputenc}
\usepackage{amsfonts}\usepackage[colorlinks,bookmarksopen,bookmarksnumbered,citecolor=blue, linkcolor=blue, urlcolor=blue]{hyperref}

\begin{document}

    \title{\bf Quantifying magic via quantum $(\alpha,\beta)$ Jensen-Shannon divergence}

    \vskip0.1in
    \author{\small Linmao Wang$^1$, Zhaoqi Wu$^1$\thanks{Corresponding author. E-mail: wuzhaoqi\_conquer@163.com}\\
        {\small\it  1. Department of Mathematics, Nanchang University,
            Nanchang 330031, P R China} }
    \date{}
    \maketitle
    \noindent {\bf Abstract} {\small }\\
    Magic states play an important role in fault-tolerant quantum
    computation, and so the quantification of magic for quantum states
    is of great significance. In this work, we propose two new magic
    quantifiers by introducing two versions of quantum $(\alpha,\beta)$
    Jensen-Shannon divergence based on the quantum $(\alpha,\beta)$
    entropy and the quantum $(\alpha,\beta)$-relative entropy,
    respectively. We derive many desirable properties for our magic
    quantifiers, and find that they are efficiently computable in
    low-dimensional Hilbert spaces. We also show that the initial
    nonstabilizerness in the input state can boost the magic generating
    power for our magic quantifiers with appropriate parameter ranges
    for a certain class of quantum gates. Our magic quantifiers may
    provide new tools for addressing some specific problems in magic
    resource theory.

    \vskip 0.1in

    \noindent {\bf Keywords:}{\small } Quantum $(\alpha,\beta)$ entropy
    $\cdot$ Quantum $(\alpha,\beta)$-relative entropy $\cdot$ Quantum
    $(\alpha,\beta)$ Jensen-Shannon divergence $\cdot$ Stabilizer
    formalism $\cdot$ Magic generating power

    \section{Introduction}\label{sec1}
    The Gottesman-Knill theorem indicates that classical computers can
    efficiently simulate any quantum computation composed solely of
    Clifford gates and preparation/measurement of stabilizer
    states\cite{Gottesman1,Gottesman2,Gottesman3,Gottesman4}.
    Nonstabilizerness, therefore, plays a crucial role in demonstrating
    quantum advantage. Veitch $et$ $al$ first introduced the stabilizer
    resource theory in quantum computation and proposed two magic
    monotones which are widely employed now\cite{Veitch1,Veitch2}. At
    the same time, several applications of magic resource theory have
    been extensively studied, such as magic state
    distillation\cite{Bravyi2,Campbell1,Campbell2,Campbell3,Dawkins,Meier,Wang},
    which has already been studied in entanglement
    theory\cite{Peres,Horodecki,Rains,Wang4,Wang5,Wang6}.

    In the past decades, magic resource theory has rapidly developed.
    Many quantifiers of magic of quantum states, for instance, the
    relative entropy of magic and the mana\cite{Veitch1}, the robustness
    of magic\cite{Howard,Gross1}, min-relative entropy and max-relative
    entropy of magic\cite{Liu}, the thauma\cite{Wang}, stabilizer
    rank\cite{Bravyi3}, stabilizer extent\cite{Heimendahl}, the
    stabilizer R\'enyi entropy\cite{Leone1,Leone2} and the $L^{p}$-norm
    magic \cite{Luo1,Luo2}, have been proposed, each of which have their
    own merits. In addition, the quantification of magic for quantum
    channels has attracted much attention in recent years. Wang
    \textit{et al}\cite{Wang} first introduced the thauma of quantum
    states and then applied the same idea to propose the thauma of
    quantum channels\cite{Wang2}. Saxena and Gour\cite{Saxena} studied
    the magic resource of multi-qubit quantum channels via the
    generalized robustness and the min-relative entropy. Seddon and
    Campbell \cite{Campbell4} presented channel robustness and magic
    capacity in $n$-qubit systems. Recently, Li and Luo\cite{Luo3}
    employed the channel-state duality and $L^{1}$-norm magic\cite{Luo2}
    to propose a new magic quantifier of channel, which is well defined
    in all dimensions.

    Quantum entropy and quantum divergence are foundational concepts
    with pervasive applications in resource theory. The concept of
    generalized quantum entropy was introduced in\cite{Hu}, while the
    unified $\left( \alpha,\beta\right) $-relative entropy was proposed
    in \cite{WangJM}, which has been applied to induce resource
    monotones in different resource theories such as coherence
    \cite{LYM} and imaginarity\cite{WuCF2025}. The uncertainty relations
    for unified $\left( \alpha,\beta\right) $-relative entropy of
    coherence have also been investigated
    extensively\cite{LYM,ChengBL2025}. The quantum Jensen-Shannon
    divergence was originated in\cite{Majtey}, and it has been proved
    that the square root of the quantum Jensen-Shannon divergence is a
    true metric on the quantum state space\cite{VD}. Moreover, the
    quantum Jensen-Shannon divergence can naturally yield monotones to
    quantify magic \cite{Tian1} and imaginarity\cite{Tian2}. In this
    work, we utilize the quantum $(\alpha,\beta)$ entropy and the quantum
    $\left( \alpha,\beta\right) $-relative entropy to define two
    versions of quantum $(\alpha,\beta)$ Jensen-Shannon divergence, and
    further propose the magic monotones induced by them.

    In magic resource theory, a natural question arises as how we can
    quantify the power of a quantum gate for generating magic resource.
    Zhu $et$ $al$\cite{Wang3} studied amortized magic in terms of the
    stabilizer R\'enyi entropy, and revealed that nonstabilizerness
    generating power can be enhanced by prior nonstabilizerness in input
    states while considering the $\alpha$-stabilizer R\'enyi entropy but
    this does not hold for the case of robustness of magic\cite{Gross1}
    or stabilizer extent\cite{Heimendahl}. When we focus on which
    quantum gate is optimal for generating magic resource, the answer
    provided by numerous studies is the $T$-gate, according to various
    magic quantifiers under certain
    conditions\cite{Tian1,Luo4,Luo5,Luo6}.

    The remainder of this work is structured as follows. In Section
    \ref{sec2}, we present a brief review of the stabilizer formalism
    and the framework of magic resource theory. In Section \ref{sec3},
    we define quantum $(\alpha,\beta)$ Jensen-Shannon divergence from
    two distinct ways and derive some properties of them. Specifically,
    we exhibit a relationship between them, from which a similar
    relationship between the corresponding magic quantifiers in Section
    4 follows. In Section \ref{sec4}, we define two magic quantifiers
    via quantum $(\alpha,\beta)$ Jensen-Shannon divergence and prove
    that our magic quantifiers are both pure-state stabilizer monotones.
    Besides, we obtain a few desirable properties and compare our magic
    quantifiers with the robustness of magic. In Section \ref{sec5}, we
    show that as in the case of the stabilizer R\'enyi entropy, the
    nonstabilizerness generating power can also be enhanced by prior
    magic in input states when considering quantum $(\alpha,\beta)$
    Jensen-Shannon divergence of magic. In Section \ref{sec6}, we give
    three detailed examples. Finally, we conclude our work with a
    summary in Section \ref{sec7}. The detailed proofs of our results
    are provided in the Appendixes.
    \section{Preliminaries}\label{sec2}
    In this section, we review the stabilizer formalism, the framework
    of magic resource theory and some basic properties of the quantum
    $(\alpha,\beta)$ entropy and the quantum $\left( \alpha,\beta\right)
    $-relative entropy.

    Let's first clarify some notations which we will use in this paper.
    Let $\mathcal{H}$ be a $d$ dimensional Hilbert space with standard
    computational basis $\left\lbrace |i\rangle\right\rbrace_{i=0}^{d-1}
    $ and $\mathcal{H}_{d^{n}}$ denotes the composite Hilbert space
    $\mathcal{H}^{\otimes n}$. Denote the set of all density operators
    by $\mathcal{D(H)}$, the set of all unitary operators by
    $\mathcal{U}(\mathcal{H})$, and the set of all stabilizer states and
    pure stabilizer states on $\mathcal{H}$ by $\mathcal{S}_{d}$ and
    $\mathcal{PS}_{d}$, respectively. We conventionally use $\rho,
    \sigma, \tau $ to represent quantum states and $|\psi \rangle,
    |\psi^{\prime} \rangle, |\phi \rangle$ to represent pure states in
    $\mathcal{D(H)}$. Moreover, $\|\cdot\|_1$ denotes the trace
    distance, i.e., $\|\rho-\sigma\|_1=\frac{1}{2}\mathrm{Tr}
    |\rho-\sigma| $.
    \subsection{Stabilizer formalism}\label{subsce2.1}
    Let $\mathbb{Z}_d$ be the ring of integers modulo $d$ and
    $\mathbb{Z}_d \times \mathbb{Z}_d$ be the direct product of
    $\mathbb{Z}_d$ and $\mathbb{Z}_d$ which can be regarded as the
    discrete phase space. The shift and boost operators are defined
    as\cite{Schwinger}
    \[
    X=\sum_{j=0}^{d-1}|j+1 \rangle \langle j|, Z=\sum_{j=0}^{d-1} \omega^{j} |j \rangle \langle j|,
    \]
    respectively, where $\omega=\text{e}^{\frac{2\pi \mathrm{i}}{d}}$. And the discrete Heisenberg-Weyl operators are defined as\cite{Appleby}
    \[
    T_{u}=\tau^{-u_{1}u_{2}}Z^{u_{1}}X^{u_{2}},
    \]
    where $\tau=\text{e}^{\frac{(d+1)\pi \mathrm{i}}{d}},u=(u_{1},u_{2}) \in \mathbb{Z}_{d} \times \mathbb{Z}_{d}$.

    The Clifford operators constitute the set defined as\cite{Veitch1}
    \begin{align*}
        \mathcal{C}_{d}=\left\lbrace U \in \mathcal{U}(\mathcal{H}): \forall u \in \mathbb{Z}_{d} \times \mathbb{Z}_{d}, \exists  u^{\prime} \in \mathbb{Z}_{d} \times \mathbb{Z}_{d}, \theta \in \mathbb{R},\, \mathrm{s.t.}\, UT_{u}U^{\dagger}=\text{e}^{\mathrm{i}\theta}T_{u^{\prime}} \right\rbrace,
    \end{align*}
    and the set of all pure stabilizer states is defined as\cite{Veitch1}
    \[
    \mathcal{PS}_{d}=\left\lbrace V|0\rangle : V \in \mathcal{C}_{d} \right\rbrace,
    \]
    while the set of all stabilizer states is the convex hull of
    $\mathcal{PS}_{d}$, i.e.\cite{Veitch1},
    \[
    \mathcal{S}_{d}=\left\lbrace \rho \in \mathcal{D}(\mathcal H):\rho=\sum_{j}p_{j}\rho_{j}, \left\lbrace \rho_{j}\right\rbrace _{j} \in \mathcal{PS}_{d}, p_{j} \geq 0, \sum_{j}p_{j}=1 \right\rbrace.
    \]
    A quantum state is a magic state if it is not a stabilizer state.

    A stabilizer operation is any map from $\rho \in
    \mathcal{D}\left(\mathcal{H}_{d^{n}}\right)$ to $\sigma \in
    \mathcal{D}\left(\mathcal{H}_{d^{m}}\right)$ composed from the
    following operations\cite{Veitch1}

    $\mathrm{(i)}$ Clifford unitaries, $\rho \rightarrow V\rho V^{\dagger}$ where $V \in \mathcal{C}_{d}$.

    $\mathrm{(ii)}$ Composition with stabilizer states, $\rho \rightarrow \rho \otimes \sigma$ where $\sigma \in \mathcal{S}_{d}$.

    $\mathrm{(iii)}$ Computational basis measurement on the final qudit, $\rho \rightarrow \frac{ \left(\mathbb{I} \otimes |i \rangle \langle i|\right) \rho  \left(\mathbb{I} \otimes |i \rangle \langle i|\right)}{\text{Tr} \left(\rho \mathbb{I} \otimes |i \rangle \langle i| \right)}$ with probability $\text{Tr} \left(\rho \mathbb{I} \otimes |i \rangle \langle i| \right)$, where $\mathbb{I}$ is the identity operator on $\mathcal{D}\left(\mathcal{H}_{d^{n-1}}\right)$.

    $\mathrm{(iv)}$ Partial trace of the final qudit, $\rho \rightarrow \text{Tr}_{n}\left(\rho\right)$.

    $\mathrm{(v)}$ The above quantum operations conditioned on the outcomes of measurements or classical randomness.

    \subsection{The framework of magic resource theory}\label{subsec2.2}
    Two important ingredients of a resource theory are free states and
    free operations. In magic resource theory, free states and free
    operations refer to stabilizer states and stabilizer operations. We first recall the framework of the quantification
    of magic resource for quantum states.

    A functional $\mathcal{M}$ that maps $\mathcal{D}(\mathcal H)$ to $[0,+\infty)$ is called a magic measure if it satisfies\textnormal{\cite{Chitambar}}

    $\mathrm{(i)}$ Faithfulness: $\mathcal{M}(\rho)=0$ if and only if $\rho \in \mathcal{S}_{d}$.

    $\mathrm{(ii)}$ Monotonicity: $\mathcal{M}(\mathcal{E}(\rho)) \leq \mathcal{M}(\rho)$ for any stabilizer operation $\mathcal{E}$.

    $\mathrm{(iii)}$ Convexity:  $\mathcal{M}(\sum_{j} p_{j} \rho_{j}) \leq \sum_{j} p_{j}\mathcal{M}(\rho_{j})$ for any  $\left\lbrace \rho_{j} \right\rbrace_{j} \in \mathcal{D}(\mathcal H)$ and probability distribution $\left\lbrace p_{j}\right\rbrace _{j}$.

    $\mathrm{(iv)}$ Strong monotonicity:  $\mathcal{M}(\rho) \geq \sum_{j} p_{j} \mathcal{M}(\rho_{j})$ for any stabilizer operation $\mathcal{E}$ satisfying $\mathcal{E}(\rho)=\sum_{j} p_{j} \rho_{j}$.

    If $\mathcal{M}$($\cdot$) only satisfies properties (i) and (ii), we say that $\mathcal{M}$($\cdot$) is a magic monotone. And $\mathcal{M}$($\cdot$) is called a pure-state stabilizer monotone if it satisfies\cite{Leone2}

    $\mathrm{(i)'}$ Faithfulness: $\mathcal{M}(|\psi \rangle \langle \psi|)=0$ if and only if $|\psi \rangle \in \mathcal{PS}_{d}$.

    $\mathrm{(ii)'}$ Monotonicity: $\mathcal{M}(|\phi \rangle \langle \phi|) \leq \mathcal{M}(|\psi \rangle \langle \psi|)$ for any stabilizer operation $\mathcal{E}$ satisfying $\mathcal{E}(|\psi \rangle \langle \psi|)=|\phi \rangle \langle \phi|$.

    A good magic quantifier should be at least a pure-state stabilizer monotone.

    \subsection{The quantum $(\alpha,\beta)$ entropy and the quantum $\left( \alpha,\beta\right) $-relative entropy}\label{subsec2.3}
    The quantum $(\alpha,\beta)$ entropy is defined as\cite{Hu}
    \[
    S_{\alpha, \beta}\left(\rho \right)=\frac{1}{(1-\alpha)\beta}\left[
    \left(\mathrm{Tr}(\rho^{\alpha})\right)^{\beta}-1\right] ,\alpha \in
    (0,1)\cup (1,+\infty),\beta \in (-\infty,0) \cup (0,+\infty),
    \]
    and the quantum $\left( \alpha,\beta\right)$-relative entropy is
    defined as\cite{WangJM}
    \[
    D_{\alpha,
        \beta}\left(\rho\|\sigma\right)=\frac{1}{(1-\alpha)\beta}\left[
    1-\left(\mathrm{Tr}\left(\rho^{\alpha}
    \sigma^{1-\alpha}\right)\right)^{\beta}\right ],\alpha \in (0,1)
    \cup (1,+\infty),\beta \in (-\infty,0) \cup (0,+\infty).
    \]

    Throughout this paper, we follow the convention that for any density operator $\rho \in \mathcal{D}(\mathcal H)$, $\rho^{-1}$ is only evaluated on its support, i.e., we retain the zero eigenvalues of $\rho^{-1}$, thus it is reasonable to define $D_{\alpha,\beta}(\rho\|\sigma)$
    for $\alpha \in (1,+\infty)$.

    The basic properties of $S_{\alpha, \beta}(\rho)$ and $D_{\alpha,
        \beta} \left(\rho\|\sigma\right)$ are summarized in the following
    two lemmas, which will be useful in the next section.
    \begin{lemma}[\!\cite{Hu,Rastegin1}] \label{lem1} Let $\left\lbrace \rho_{j}\right\rbrace _{j} \in \mathcal{D}(\mathcal H)$, and $\left\{p_{j} \right\}_{j}$ be a probability distribution. Then we have

        $\mathrm{(i)}$ (Concavity) For $\alpha \in (0,1),\beta \in (-\infty,0)\cup (0,1]$, it holds that
        \[
        S_{\alpha, \beta}\left(\sum_{j}^{}p_{j} \rho_{j}\right) \geq \sum_{j}^{} p_{j}S_{\alpha, \beta}\left(\rho_{j} \right),
        \]
        and for $\alpha \in (1,+\infty),\beta \in [1,+\infty) $, it holds that
        \[
        \sum_{j}^{} p_{j}S_{\alpha, \beta}(\rho_{j} )\leq S_{\alpha, \beta}\left(\sum_{j}^{}p_{j} \rho_{j}\right) \leq \sum_{j}^{} p_{j}S_{\alpha, \beta}\left(\rho_{j}\right)+F_{\alpha}^{\beta}\left( \left\lbrace p_{j}\right\rbrace _{j}\right),
        \]
        where
        \[
        F_{\alpha}^{\beta}\left( \left\lbrace p_{j}\right\rbrace_{j} \right)=\frac{1}{(1-\alpha)\beta}\left( \sum_{j}^{}p_{j}^{\alpha\beta}-1 \right).
        \]

        $\mathrm{(ii)}$ For $\alpha \in (0,1)\cup (1,+\infty),\beta \in (-\infty,0) \cup (0,+\infty)$, it holds that

        \[
        S_{\alpha, \beta}\left(\rho \otimes\sigma\right)=S_{\alpha, \beta}\left(\rho \right)+S_{\alpha, \beta}\left(\sigma \right)+(1-\alpha)\beta S_{\alpha, \beta}\left(\rho \right) S_{\alpha, \beta}\left(\sigma \right).
        \]

        $\mathrm{(iii)}$ (Unitary invariance) For $\alpha \in (0,1)\cup (1,+\infty),\beta \in (-\infty,0) \cup (0,+\infty)$ and any unitary operator $U \in \mathcal{U}(\mathcal{H})$, it holds that
        \[
        S_{\alpha, \beta}\left(\rho\right) =S_{\alpha, \beta}\left(U \rho U^{\dagger}\right).
        \]

        $\mathrm{(iv)}$ (Lipschitz continuity) For $\alpha \in (1,+\infty) , \beta \in [1,+\infty)$, it holds that
        \[
        |S_{\alpha,\beta}(\rho)-S_{\alpha,\beta}(\sigma)| \leq \frac{\alpha}{\alpha-1}\|\rho-\sigma\|_{1}.
        \]
    \end{lemma}
    \begin{lemma}[\!\cite{WangJM}]\label{lem2} Let $\left\lbrace \rho_{j}\right\rbrace _{j} ,\left\lbrace \sigma_{j}\right\rbrace _{j} \in \mathcal{D}(\mathcal H)$, and $\left\{p_{j} \right\}_{j}$ be a probability distribution. Then we have

        $\mathrm{(i)}$ For $\alpha \in (0,1)\cup (1,+\infty),\beta \in (-\infty,0) \cup (0,+\infty)$, it holds that
        \[
        D_{\alpha, \beta}\left(\rho\|\sigma\right) \geq 0,
        \]
        and the equality holds if and only if $\rho=\sigma$.

        $\mathrm{(ii)}$ For $\alpha \in (0,1)\cup (1,+\infty),\beta \in (-\infty,0) \cup (0,+\infty)$, it holds that
        \[
        D_{\alpha, \beta}\left(\rho_{1}\otimes \rho_{2}\|\sigma_{1}\otimes \sigma_{2}\right)=D_{\alpha, \beta}\left(\rho_{1}\|\sigma_{1}\right)+D_{\alpha, \beta}\left(\rho_{2}\|\sigma_{2}\right)+(\alpha-1)\beta D_{\alpha, \beta}\left(\rho_{1}\|\sigma_{1}\right)D_{\alpha, \beta}\left(\rho_{2}\|\sigma_{2}\right).
        \]

        $\mathrm{(iii)}$ (Monotonicity) For any quantum operation $\mathcal{E}$ and $\alpha \in (0,1) , \beta \in (-\infty,0)\cup (0,1]$, it holds that
        \[
        D_{\alpha, \beta}\left(\mathcal{E}(\rho)\|\mathcal{E}(\sigma)\right) \leq D_{\alpha, \beta}\left(\rho\|\sigma\right).
        \]

        $\mathrm{(iv)}$ (Unitary invariance) For $\alpha \in (0,1)\cup (1,+\infty),\beta \in (-\infty,0) \cup (0,+\infty)$ and any unitary operator $U \in \mathcal{U}(\mathcal{H})$, it holds that
        \[
        D_{\alpha, \beta}\left(U\rho U^{\dagger}\|U \sigma U^{\dagger} \right)=D_{\alpha,
            \beta}\left(\rho\|\sigma\right).
        \]

        $\mathrm{(v)}$ (Joint convexity) For $\alpha \in (0,1) , \beta \in (-\infty,0)\cup (0,1]$, it holds that
        \[
        D_{\alpha, \beta} \left( \sum_{j}^{} p_{j} \rho_{j} \Big\| \sum_{j}^{} p_{j} \sigma_{j}\right) \leq \sum_{j}^{} p_{j} D_{\alpha, \beta} \left(\rho_{j},\sigma_{j}\right).
        \]
    \end{lemma}
    \section{Quantum $(\alpha,\beta)$ Jensen-Shannon divergence}\label{sec3}
    In this section, we present the definition and some properties of the quantum $(\alpha,\beta)$ Jensen-Shannon divergence.

    The quantum Jensen-Shannon divergence is defined as\cite{Majtey}
    \[
    J(\rho,\sigma) = \frac{1}{2}\left[D\left(\rho\Big\|\frac{\rho+\sigma}{2}\right) + D\left(\sigma\Big\|\frac{\rho+\sigma}{2}\right)\right] = S\left(\frac{\rho+\sigma}{2}\right) - \frac{1}{2}S(\rho) -
    \frac{1}{2}S(\sigma),
    \]
    where $S(\rho)=-\mathrm{Tr}\rho\log\rho$ is the von Neumann entropy
    of $\rho$ and
    $D(\rho\|\sigma)=\mathrm{Tr}(\rho\log\rho-\rho\log\sigma)$ is the
    relative entropy between $\rho$ and $\sigma$.

    Motivated by the above concepts, we now define the quantum $(\alpha,\beta)$ Jensen-Shannon divergence as follows.
    \begin{definition} \label{def1} For $\alpha \in (0,1)\cup (1,+\infty),\beta \in (-\infty,0) \cup (0,+\infty)$, we define two kinds of quantum $(\alpha,\beta)$ Jensen-Shannon divergence as
        \begin{equation}\label{eq1}
            \text{J}_{\alpha, \beta}(\rho, \sigma) = S_{\alpha, \beta}\left(\frac{\rho+\sigma}{2}\right)-\frac{1}{2}S_{\alpha, \beta}\left(\rho \right) -\frac{1}{2}S_{\alpha, \beta}\left(\sigma \right),
        \end{equation}
        and
        \begin{equation}\label{eq2}
            J_{\alpha, \beta}^{\prime}(\rho, \sigma)=\frac{1}{2}\left[D_{\alpha, \beta} \left(\rho\Big\|\frac{\rho+\sigma}{2}\right)+D_{\alpha, \beta} \left(\sigma\Big\|\frac{\rho+\sigma}{2}\right)\right].
        \end{equation}
    \end{definition}
    \begin{remark} \label{rem1} Eqs. $\mathrm{(\ref{eq1})}$ and $\mathrm{(\ref{eq2})}$ both degenerate to the quantum Jensen-Shannon divergence introduced in \cite{Majtey} when $\alpha \rightarrow 1$, since $S_{\alpha,\beta}(\rho)$ and $D_{\alpha,\beta}(\rho\|\sigma)$ degenerates to $S(\rho)$ and $D(\rho\|\sigma)$ respectively when $\alpha\rightarrow
        1$. Moreover, Eqs. $\mathrm{(\ref{eq1})}$ and $\mathrm{(\ref{eq2})}$
        reduce to two versions of quantum Jensen-Tsallis divergence proposed
        in \cite{Sra2021LAA} when $\beta=1$, which do not coincide.
    \end{remark}
    \begin{proposition} \label{pro1} For any $\alpha \in (0,1) \cup (1,2),\beta \in (-\infty,0) \cup (0,+\infty)$, it holds that
        \begin{equation}\label{eq3}
            J_{\alpha, \beta}\left( |\psi \rangle \langle \psi|,|\phi \rangle \langle \phi|\right)=J_{2-\alpha, \beta}^{\prime} \left( |\psi \rangle \langle \psi|,|\phi \rangle \langle \phi|\right),
        \end{equation}
    \end{proposition}
    We leave the proof of Proposition \ref{pro1} in Appendix \ref{A.1.}.

    Based on Lemma \ref{lem1} and Lemma \ref{lem2}, the properties of $J_{\alpha,\beta}(\rho,\sigma)$ and $J_{\alpha,\beta}^{\prime}(\rho,\sigma)$ can be derived as follows.
    \begin{theorem} \label{thm1} Let $\rho_{1},\rho_{2},\sigma_{1},\sigma_{2} \in \mathcal{D}(\mathcal H)$, $\left\{|\psi_{j} \rangle\right\}_{j}$, $\left\{|\phi_{j} \rangle\right\}_{j}$ are pure states on $\mathcal{H}$, and $\left\{p_{j} \right\}_{j}$ is a probability distribution. Then we have

        $\mathrm{(i)}$ It holds that
        \[
        J_{\alpha, \beta}(\rho, \sigma)\geq 0
        \]
        for $\alpha \in (0,1),\beta \in (-\infty,0)\cup (0,1)$ or $\alpha \in (1,+\infty),\beta \in [1,+\infty) $,
        and
        \[
        J_{\alpha, \beta}(\rho, \sigma) \leq \frac{1}{(1-\alpha)\beta}\left(2^{1-\alpha\beta}-1\right)
        \]
        for $\alpha \in (1,+\infty),\beta \in [1,+\infty) $.

        $\mathrm{(ii)}$ For $\alpha \in (0,1)\cup (1,+\infty),\beta \in (-\infty,0) \cup (0,+\infty)$, it holds that
        \[
        J_{\alpha, \beta}(\rho \otimes \tau, \sigma \otimes \tau)=\left[ 1+(1-\alpha)\beta S_{\alpha, \beta}(\tau)\right]J_{\alpha, \beta}(\rho, \sigma).
        \]
        Specifically, when $\tau$ is a pure state, we have $J_{\alpha,\beta}(\rho \otimes \tau, \sigma \otimes \tau)=J_{\alpha,
            \beta}(\rho,\sigma)$.

        $\mathrm{(iii)}$ (Unitary invariance) For $\alpha \in (0,1)\cup (1,+\infty),\beta \in (-\infty,0) \cup (0,+\infty)$ and any unitary operator $U  \in \mathcal{U}(\mathcal{H})$, it holds that
        \[
        J_{\alpha, \beta}(U\rho U^{\dagger}, U\sigma U^{\dagger})=J_{\alpha, \beta}(\rho, \sigma).
        \]

        $\mathrm{(iv)}$ (Symmetry) For $\alpha \in (0,1)\cup (1,+\infty),\beta \in (-\infty,0) \cup (0,+\infty)$, it holds that
        \[
        J_{\alpha, \beta}(\rho, \sigma) =J_{\alpha, \beta}(\sigma, \rho).
        \]

        $\mathrm{(v)}$ (Lipschitz continuity) For $\alpha \in (1,+\infty) , \beta \in [1,+\infty)$, $J_{\alpha,\beta}(\rho,\sigma)$ is Lipschitz continuous in the first or second entry, i.e.,
        \[
        |J_{\alpha,\beta}(\rho_{1},\sigma)-J_{\alpha,\beta}(\rho_{2},\sigma)| \leq \frac{\alpha}{\alpha-1}\|\rho_{1}-\rho_{2}\|_{1},
        \]
        and
        \[
        |J_{\alpha,\beta}(\rho,\sigma_{1})-J_{\alpha,\beta}(\rho,\sigma_{2})| \leq \frac{\alpha}{\alpha-1}\|\sigma_{1}-\sigma_{2}\|_{1}.
        \]

    \end{theorem}
    We leave the proof of Theorem \ref{thm1} in Appendix \ref{A.2.}.

    \begin{theorem} \label{thm2} Let $\left\lbrace \rho_{j}\right\rbrace _{j}, \left\lbrace \sigma_{j}\right\rbrace _{j} \in \mathcal{D}(\mathcal H)$, and $\left\{p_{j} \right\}_{j}$ be a probability distribution. Then we have

        $\mathrm{(i)}$  For $\alpha \in (0,1)\cup (1,+\infty),\beta \in (-\infty,0) \cup (0,+\infty)$, it holds that
        \[
        J_{\alpha, \beta}^{\prime}(\rho, \sigma) \geq 0,
        \]
        and the equality holds if and only if $\rho=\sigma$.

        $\mathrm{(ii)}$  For $\alpha \in (0,1)\cup (1,+\infty),\beta \in (-\infty,0) \cup (0,+\infty)$, it holds that
        \[
        J_{\alpha, \beta}^{\prime}(\rho \otimes \tau, \sigma \otimes \tau)=J_{\alpha, \beta}^{\prime}(\rho, \sigma).
        \]

        $\mathrm{(iii)}$ (Monotonicity) For any quantum operation $\mathcal{E}$ and $\alpha \in (0,1) , \beta \in (-\infty,0)\cup (0,1]$, it holds that
        \[
        J_{\alpha, \beta}^{\prime} \left( \mathcal{E}(\rho),\mathcal{E}(\sigma)\right) \leq J_{\alpha, \beta}^{\prime}(\rho, \sigma).
        \]

        $\mathrm{(iv)}$ (Unitary invariance) For $\alpha \in (0,1)\cup (1,+\infty),\beta \in (-\infty,0) \cup (0,+\infty)$ and any unitary operator  $U \in \mathcal{U}(\mathcal{H})$, it holds that
        \[
        J_{\alpha, \beta}^{\prime}(U\rho U^{\dagger}, U\sigma U^{\dagger})=J_{\alpha, \beta}^{\prime}(\rho, \sigma).
        \]

        $\mathrm{(v)}$ (Joint convexity) For $\alpha \in (0,1) , \beta \in (-\infty,0)\cup (0,1]$, it holds that
        \[
        J_{\alpha, \beta}^{\prime} \left( \sum_{j}^{} p_{j} \rho_{j},\sum_{j}^{} p_{j} \sigma_{j}\right) \leq \sum_{j}^{} p_{j}J_{\alpha, \beta}^{\prime} \left(\rho_{j},\sigma_{j}\right).
        \]

        $\mathrm{(vi)}$ (Symmetry) For $\alpha \in (0,1)\cup (1,\infty),\beta \in (-\infty,0) \cup (0,+\infty)$, it holds that
        \[
        J_{\alpha, \beta}^{\prime}(\rho, \sigma)=J_{\alpha, \beta}^{\prime}(\sigma, \rho).
        \]
    \end{theorem}
    We leave the proof of Theorem \ref{thm2} in Appendix \ref{A.3.}.

    \begin{remark}\label{rem2} By applying Proposition $\mathrm{\ref{pro1}}$ and Theorem $\mathrm{\ref{thm2}}$ $\mathrm{(iii)}$, for $\alpha \in (1,2), \beta \in (-\infty,0)\cup (0,1]$ and any stabilizer operation $\mathcal{E}$ satisfying $\mathcal{E}(|\psi \rangle \langle \psi|)=|\psi^{\prime} \rangle \langle \psi^{\prime}|$ which means that $\mathcal{E}$ maps a pure state to another pure state, we have
        \begin{equation}\label{eq4}
            J_{\alpha,\beta} \left( \mathcal{E}\left( |\psi \rangle \langle \psi|\right) ,\mathcal{E}(|\phi \rangle \langle \phi|)\right)  \leq J_{\alpha, \beta} \left( |\psi \rangle \langle \psi|,|\phi \rangle \langle
            \phi|\right).
        \end{equation}

    \end{remark}
    \section{Quantum $(\alpha,\beta)$ Jensen-Shannon divergence of magic}\label{sec4}
    In this section, we present the definition and some properties of
    the quantum $(\alpha,\beta)$ Jensen-Shannon divergence of magic, and
    compare our magic quantifiers with other existing ones.
    \subsection{Two new stabilizer monotones}\label{subsec4.1}
    We start by defining two magic quantifiers for pure states via the quantum $(\alpha,\beta)$ Jensen-Shannon divergence.
    \begin{definition}\label{def2} For $\alpha \in (0,1)\cup (1,+\infty),\beta \in (-\infty,0) \cup (0,+\infty)$, define
        \begin{equation}\label{eq5}
            M_{\alpha, \beta}  \left( |\psi \rangle \langle \psi|\right) =\min_{|\phi \rangle \in \mathcal{PS}_{d}} J_{\alpha, \beta}\left( |\psi \rangle \langle \psi|,|\phi \rangle \langle \phi|\right),
        \end{equation}
        \begin{equation}\label{eq6}
            m_{\alpha, \beta} \left( |\psi \rangle \langle \psi|\right) =\min_{|\phi \rangle \in \mathcal{PS}_{d}}J_{\alpha, \beta}^{\prime}\left( |\psi \rangle \langle \psi|,|\phi \rangle \langle \phi|\right).
        \end{equation}
    \end{definition}
    Next we use convex roof construction to define the quantum $(\alpha,\beta)$ Jensen-Shannon divergence of magic for mixed states.
    \begin{definition}\label{def3} For $\alpha \in (0,1)\cup (1,+\infty),\beta \in (-\infty,0) \cup (0,+\infty)$, we define two kinds of  quantum $(\alpha,\beta)$ Jensen-Shannon divergence of magic as
        \begin{equation}\label{eq7}
            M_{\alpha, \beta} \left(\rho\right)=\min_{\left\lbrace (p_j,|\psi_{j} \rangle)\right\rbrace }\sum_{j}^{}p_{j}M_{\alpha, \beta}  \left( |\psi_{j} \rangle \langle \psi_{j}|\right),
        \end{equation}
        \begin{equation}\label{eq8}
            m_{\alpha, \beta} \left(\rho\right)=\min_{\left\lbrace (p_j,|\psi_{j} \rangle)\right\rbrace }\sum_{j}^{}p_{j}m_{\alpha, \beta}  \left( |\psi_{j} \rangle \langle \psi_{j}|\right),
        \end{equation}
        where $\rho=\sum_{j}^{}p_{j} |\psi_{j} \rangle \langle \psi_{j}| $ are all pure-state decompositions of $\rho$.
    \end{definition}
    From the proof of Proposition \ref{pro1}, we can see that
    \begin{equation}\label{eq9}
        M_{\alpha, \beta}  \left( |\psi \rangle \langle \psi|\right) =\min_{|\phi \rangle \in \mathcal{PS}_{d}}\frac{1}{(1-\alpha)\beta}\left[ \left(\lambda_{1}^{\alpha}+\lambda_{2}^{\alpha}\right)^{\beta}-1\right],
    \end{equation}
    and
    \begin{equation}\label{eq10}
        m_{\alpha, \beta}  \left( |\psi \rangle \langle \psi|\right) =\min_{|\phi \rangle \in \mathcal{PS}_{d}} \frac{1}{(1-\alpha)\beta} \left[ 1- \left(\lambda_{1}^{2-\alpha}+\lambda_{2}^{2-\alpha}\right)^{\beta}\right],
    \end{equation}
    where $ \lambda_1=\frac{1+|\langle \psi|\phi
        \rangle|}{2}$ and $\lambda_2=\frac{1-|\langle \psi|\phi
        \rangle|}{2}$ are the eigenvalues of $\frac{|\psi \rangle \langle
        \psi|+|\phi \rangle \langle \phi|}{2}$. This implies that
    \begin{equation}\label{eq11}
        m_{\alpha, \beta}  \left(\rho\right)= M_{2-\alpha, \beta}  \left(\rho \right),
    \end{equation}
    for $\alpha \in (0,1) \cup (1,2),\beta \in (-\infty,0) \cup (0,+\infty)$.
    \begin{remark}\label{rem3} Both of the quantities in Eqs. $\mathrm{(\ref{eq7})}$ and $\mathrm{(\ref{eq8})}$ degenerate to the quantum Jensen-Shannon divergence of magic in \textnormal{\cite{Tian1}} when $\alpha \rightarrow 1$.
    \end{remark}
    Next we give some properties of $M_{\alpha, \beta} \left(\rho\right)$ and $m_{\alpha, \beta} \left(\rho\right)$.
    \begin{theorem}\label{thm3} Let $\left\lbrace \rho_{j}\right\rbrace _{j} \in \mathcal{D}(\mathcal H) $, and $\left\{p_{j} \right\}_{j}$ be a probability distribution. Then we have

        $\mathrm{(i)}$ (Faithfulness) For $\alpha \in (0,1)\cup (1,+\infty),\beta \in (-\infty,0) \cup (0,+\infty)$, it holds that
        \[
        M_{\alpha, \beta} \left(\rho\right) \geq 0,
        \]
        and the equality holds if and only if $\rho \in \mathcal{S}_{d}$.

        $\mathrm{(ii)}$ (Invariance under Clifford operations) For any Clifford operator $V \in \mathcal{C}_{d}$ and   $\alpha \in (0,1)\cup (1,+\infty),\beta \in (-\infty,0) \cup (0,+\infty)$, it holds that
        \[
        M_{\alpha, \beta} \left(V\rho V^{\dagger}\right)=M_{\alpha, \beta} \left(\rho\right).
        \]

        $\mathrm{(iii)}$ (Convexity) For $\alpha \in (0,1)\cup (1,+\infty),\beta \in (-\infty,0) \cup (0,+\infty)$, it holds that
        \[
        M_{\alpha, \beta} \left(\sum_{j}^{} p_{j} \rho_{j}\right) \leq \sum_{j}^{} p_{j}M_{\alpha, \beta} \left(\rho_{j}\right).
        \]

        $\mathrm{(iv)}$ (Monotonicity) For $\alpha \in (1,2), \beta \in (-\infty,0)\cup (0,1]$ and any stabilizer operation $\mathcal{E}$ which obeys $\mathcal{E}(|\psi \rangle \langle \psi|)=|\phi \rangle \langle \phi|$, it holds that
        \[
        M_{\alpha, \beta} \left(|\phi \rangle \langle \phi| \right) \leq M_{\alpha, \beta}  \left(|\psi \rangle \langle \psi|\right).
        \]

        $\mathrm{(v)}$ For $\alpha \in (1,2), \beta \in (-\infty,0)\cup (0,1]$, it holds that
        \[
        M_{\alpha, \beta}  \left(\rho \otimes \sigma\right) \geq M_{\alpha, \beta}  \left(\rho\right),
        \]
        and the equality holds if $\sigma \in \mathcal{S}_{d}$.

        $\mathrm{(vi)}$ (Lipschitz continuity) For $\alpha \in (1,+\infty) , \beta \in [1,+\infty)$, $M_{\alpha,\beta}(\rho)$ is Lipschitz continuous for pure states, i.e.,
        \[
        |M_{\alpha,\beta}(|\psi_{1} \rangle \langle \psi_{1}|)-M_{\alpha,\beta}(|\psi_{2} \rangle \langle \psi_{2}|)| \leq \frac{\alpha}{\alpha-1} || |\psi_{1} \rangle \langle \psi_{1}|-|\psi_{2} \rangle \langle \psi_{2}| ||_{1}.
        \]
    \end{theorem}
    We leave the proof of Theorem \ref{thm3} in Appendix \ref{A.4.}.

    By imitating the proof of Theorem \ref{thm3} and utilizing Theorem \ref{thm2}, we can also prove that $m_{\alpha,\beta}$ exhibit the following desirable properties.
    \begin{theorem}\label{thm4} Let $\left\lbrace \rho_{j}\right\rbrace _{j} \in \mathcal{D}(\mathcal H)$, and $\left\{p_{j} \right\}_{j}$ be a probability distribution. Then we have

        $\mathrm{(i)}$ (Faithfulness) For $\alpha \in (0,1)\cup (1,+\infty),\beta \in (-\infty,0) \cup (0,+\infty)$, it holds that
        \[
        m_{\alpha, \beta} \left(\rho\right) \geq 0,
        \]
        and the equality holds if and only if  $\rho \in \mathcal{S}_{d}$.

        $\mathrm{(ii)}$ (Invariance under Clifford operations) For any Clifford operator $V \in \mathcal{C}_{d}$ and  $\alpha \in (0,1)\cup (1,+\infty),\beta \in (-\infty,0) \cup (0,+\infty)$, it holds that
        \[
        m_{\alpha, \beta} \left(V\rho V^{\dagger}\right)=m_{\alpha, \beta} \left(\rho\right).
        \]

        $\mathrm{(iii)}$  (Convexity) For $\alpha \in (0,1)\cup (1,+\infty),\beta \in (-\infty,0) \cup (0,+\infty)$  and $\rho=\sum_{j}p_{j}\rho_{j}$, convexity holds, i.e.,
        \[
        m_{\alpha, \beta} \left(\sum_{j}^{} p_{j} \rho_{j}\right) \leq \sum_{j}^{} p_{j}m_{\alpha, \beta} \left(\rho_{j}\right).
        \]

        $\mathrm{(iv)}$ (Monotonicity) For $\alpha \in (0,1) , \beta \in (-\infty,0)\cup (0,1]$ and any stabilizer operation $\mathcal{E}$ which satisfying $\mathcal{E}(|\psi \rangle \langle \psi|)=|\phi \rangle \langle \phi|$, it holds that
        \[
        m_{\alpha, \beta} \left(|\phi \rangle \langle \phi|  \right) \leq m_{\alpha, \beta}  \left(|\psi \rangle \langle \psi|\right).
        \]

        $\mathrm{(v)}$ For $\alpha \in (0,1) , \beta \in (-\infty,0)\cup (0,1]$, it holds that
        \[
        m_{\alpha, \beta}  \left(\rho \otimes \sigma\right) \geq m_{\alpha, \beta}  \left(\rho\right),
        \]
        and the equality holds if $\sigma \in \mathcal{S}_{d}$.
    \end{theorem}

    The above two theorems show that the two versions of quantum $(\alpha,\beta)$ Jensen-Shannon divergence of magic are both pure-state stabilizer monotones with suitable parameter ranges which will be very helpful while dealing with some pure-state issues. Since $M_{\alpha,\beta}(\rho)$ and $m_{\alpha,\beta}(\rho)$ share similar properties and are closely related, we may only focus on discussing $M_{\alpha,\beta}(\rho)$ in the rest of our work.

    We first present two obvious lemmas which are easy to verify.
    \begin{lemma}\label{lem3} The function
        \begin{equation}\label{eq12}
            g(x,\alpha,\beta)=\frac{x^{\beta}-1}{(1-\alpha)\beta}
        \end{equation}
        satisfies

        $\mathrm{(i)} $ $g(x,\alpha,\beta)$ is strictly monotonically increasing for $\alpha \in (0,1) ,\beta \in (-\infty,0) \cup (0,+\infty)$ and strictly monotonically decreasing for $\alpha \in (1,+\infty) ,\beta \in (-\infty,0) \cup (0,+\infty)$ with respect to $x\in
        (0,+\infty)$;

        $ \mathrm{(ii)}$ $g(x,\alpha,\beta)$ is strictly convex with respect to $x\in (0,+\infty)$ for $\alpha \in (1,+\infty), \beta \in (-\infty,0) \cup (0,1)$.
    \end{lemma}
    \begin{lemma}\label{lem4} The function
        \begin{equation}\label{eq13}
            f(\lambda,\alpha)= \lambda^{\alpha}+(1-\lambda)^{\alpha}
        \end{equation}
        is monotonically decreasing with respect to $\lambda \in (\frac{1}{2},1) $ for $\alpha \in (0,1)$ and monotonically increasing  with respect to $\lambda \in (\frac{1}{2},1) $  for $\alpha \in (1,+\infty)$.
    \end{lemma}
    %$(ii)$ $f(\lambda,\alpha)$ is convex function respect to $\alpha \in (1,+\infty)$ with $\lambda \in (\frac{1}{2},1)$.

    It seems a little difficult to calculate our magic monotones even
    for pure states. However, we can simplify the calculation of Eq.
    (\ref{eq5}) by means of the following proposition.
    \begin{proposition}\label{pro2} For $\alpha \in (0,1)\cup (1,+\infty) , \beta \in(-\infty,0)\cup (0,+\infty)$, we have
        \begin{equation}\label{eq14}
            M_{\alpha,\beta} \left(|\psi \rangle \langle \psi|\right)= \frac{1}{(1-\alpha)\beta}\left\lbrace \left[  \left(\frac{1+|c_{\psi}|}{2} \right)^{\alpha}+\left(\frac{1-|c_{\psi}|}{2} \right) ^{\alpha}    \right] ^{\beta}-1\right\rbrace,
        \end{equation}
        where $|c_{\psi}|=\max\limits_{|\phi\rangle \in \mathcal{PS}_{d}}|\langle \psi | \phi\rangle|$.
    \end{proposition}
    We leave the proof of Proposition \ref{pro2} in Appendix \ref{A.5.}.
    From the above proposition, we can directly determine $M_{\alpha, \beta}  \left( |\psi \rangle \langle \psi |\right)$ by calculating $|c_{\psi}|$, which may be easily calculated for any pure state $|\psi\rangle$ in qubit or qutrit systems.

    \subsection{Relationship with other magic quantifiers}\label{subsec4.2}

    In this subsection, we compare the relationship between our magic quantifier $M_{\alpha,\beta}$ with other existing magic quantifiers.

    The robustness of magic of $\rho$ is defined as\cite{Gross1}
    \begin{equation}\label{eq15}
        R(\rho)= \inf_{\sigma \in \mathcal{S}_{d}} \left\{ s \geq 0 \middle | \frac{\rho+s\sigma}{1+s} \in \mathcal{S}_{d}  \right\},
    \end{equation}
    and the min-relative entropy of magic is defined by \cite{Liu}
    \begin{equation}\label{eq16}
        D_{\mathrm{min}}(\rho)=\min_{\sigma \in \mathcal{S}_{d} }\left\lbrace -\mathrm{log}\left[\mathrm{Tr}\left( \mathcal{\pi}_{\rho} \sigma\right) \right]\right\rbrace,
    \end{equation}
    where $\mathcal{\pi}_{\rho} $ is the projector onto the support of $\rho$. These two quantifiers exhibit the following relationship\cite{Liu}
    \begin{equation}\label{eq17}
        D_{\mathrm{min}}(\rho) \leq \log\left( 1+R(\rho)\right).
    \end{equation}

    When $\rho=|\psi\rangle \langle \psi|$ is a pure state, we have $D_{\mathrm{min}} \left(|\psi \rangle \langle \psi |\right) = -\mathrm{log} F\left(|\psi \rangle \langle \psi |\right)$, where
    \begin{equation}\label{eq18}
        F\left(|\psi \rangle \langle \psi |\right) =\max_{ |\phi \rangle \in \mathcal{PS}_{d}} |\langle \psi | \phi \rangle|^{2}
    \end{equation}
    is the stabilizer fidelity\cite{Bravyi}, which implies that
    \begin{equation}\label{eq19}
        \frac{1}{1+R\left(|\psi \rangle \langle \psi |\right)} \leq F\left(|\psi \rangle \langle \psi |\right).
    \end{equation}

    In view of the monotonicity of classical Tsallis $\alpha$-entropy with respect to $\alpha \in  (1,+\infty)$ and the monotonicity of classical unified $\left(\alpha,\beta\right)$-entropy with respect to $\beta  \in (1,+\infty)$, we have the following two lemmas.
    \begin{lemma}\label{lem5} For $\lambda \in [0,1]$,
        \begin{equation}\label{eq20}
            H_{\alpha} \left\{\left(\lambda,1-\lambda\right)\right\}=\frac{1}{1-\alpha}\left[\lambda^{\alpha}+\left(1-\lambda\right)^{\alpha}-1\right]
        \end{equation}
        is monotonically decreasing with respect to $\alpha \in (1,+\infty)$ and monotonically increasing with respect to $\alpha \in (0,1)$.
    \end{lemma}
    \begin{lemma}\label{lem6} For $\lambda \in [0,1]$,
        \begin{equation}\label{eq21}
            H_{\alpha,\beta} \left\{\left(\lambda,1-\lambda\right)\right\}=\frac{1}{(1-\alpha)\beta}\left\{\left[\lambda^{\alpha}+\left(1-\lambda\right)^{\alpha}\right]^{\beta}-1\right\}
        \end{equation}
        is monotonically decreasing in $\alpha \in (1,+\infty)$ and monotonically increasing in $\alpha \in (0,1)$ with respect to $\beta \in (1,+\infty)$.
    \end{lemma}
    The relationship between the quantum $(\alpha,\beta)$ Jensen-Shannon divergence of magic and the robustness of magic can be established as follows.
    \begin{proposition}\label{pro3} For $\alpha \in (0,1) \cup (1,+\infty),\lambda \in [1/2,1], \beta  \in (1,+\infty)$, it holds that
        \begin{equation}\label{eq22}
            M_{\alpha,\beta}\left(|\psi \rangle \langle \psi |\right) \leq t_{0}-  \frac{1}{1+R\left(|\psi \rangle \langle \psi |\right)},
        \end{equation}
        where
        \[
        t_{0}=(2\lambda_{0}-1)^{2}+H \left\{\left(\lambda_{0},1-\lambda_{0}\right)\right\},
        \]
        and $\lambda_{0}$ is the solution of equation  $\lambda=(1-\lambda)16^{2\lambda-1}$.
    \end{proposition}
    We leave the proof of Proposition \ref{pro3} in Appendix \ref{A.6.}.
    \section{Magic generating power of  quantum gates via quantum $(\alpha,\beta)$ Jensen–Shannon divergence}\label{sec5}
    In this section, we discuss magic generating power of quantum gates via quantum $(\alpha,\beta)$ Jensen-Shannon divergence.

    For any $U \in \mathcal{U}(\mathcal{H})$, we define the magic generating power of $U$ based on $M_{\alpha,\beta}$ as
    \begin{equation}\label{eq23}
        \mathcal{M}_{\alpha,\beta}(U)=\max_{ \rho \in \mathcal{S}_{d}}M_{\alpha,\beta}(U\rho U^{\dagger}).
    \end{equation}

    $\mathcal{M}_{\alpha,\beta}(U)$ have the following properties.
    \begin{theorem}\label{thm5} For any $U,U_{1},U_{2} \in \mathcal{U}(\mathcal{H})$, we have

        $\mathrm{(i)}$ For $\alpha \in (0,1) \cup (1,+\infty), \beta \in (-\infty,0) \cup(0,+\infty)$, it holds that
        \[
        \mathcal{M}_{\alpha,\beta}(U) \geq 0,
        \]
        and equality holds if and only if $U \in \mathcal{C}_{d}$.

        $\mathrm{(ii)}$ For $\alpha \in (0,1) \cup (1,+\infty), \beta \in (-\infty ,0)\cup(0,+\infty)$ and any $V_{1},V_{2} \in \mathcal{C}_{d}$, it holds that
        \[
        \mathcal{M}_{\alpha,\beta}(U)=M_{\alpha,\beta}(V_{1}UV_{2}).
        \]

        $\mathrm{(iii)}$ For $\alpha \in (1,2), \beta \in (-\infty,0)\cup
        (0,1]$, it holds that
        \[
        \mathcal{M}_{\alpha,\beta}(U_{1} \otimes U_{2}) \geq \mathcal{M}_{\alpha,\beta}(U_{1}).
        \]
    \end{theorem}
    We leave the proof of Theorem \ref{thm5} in Appendix \ref{A.7.}.

    Thus we can also simplify the calculation of Eq. $(\ref{eq23})$ by the following proposition.
    \begin{proposition}\label{pro4} For $\alpha \in (0,1) \cup (1,+\infty), \beta \in(-\infty,0)\cup (0,+\infty)$ and any $U \in \mathcal{U}(\mathcal{H})$, we have
        \begin{equation}\label{eq24}
            \mathcal{M}_{\alpha,\beta}(U)=\frac{1}{(1-\alpha)\beta}\left\lbrace \left[  \left(\frac{1+|C_{U}|}{2} \right)^{\alpha}+\left(\frac{1-|C_{U}|}{2} \right) ^{\alpha}    \right] ^{\beta}-1\right\rbrace,
        \end{equation}
        where
        \begin{equation}\label{eq25}
            |C_{U}|=\min_{|\phi\rangle \in \mathcal{PS}_{d}}\max_{ |\psi \rangle \in \mathcal{PS}_{d}}|\langle \psi |U|\phi \rangle |.
        \end{equation}
    \end{proposition}
    We leave the proof of Proposition \ref{pro4} in Appendix \ref{A.8.}.
    Proposition \ref{pro4} reveals that $\mathcal{M}_{\alpha,\beta}(U)$ is totally determined by $|C_{U}|$.

    By proving Lemma \ref{lem7}, we can apply Proposition \ref{pro4} to give Proposition \ref{pro5}.
    \begin{lemma}\label{lem7} The function
        \begin{equation}\label{eq26}
            w(x,\alpha)=\mathrm{cos}^{2\alpha}x+\mathrm{sin}^{2\alpha}x
        \end{equation}
        satisfies

        $\mathrm{(i)}$ $w(x,\alpha)$ is strictly monotonically decreasing for $\alpha \in (1,2)$ with respect to $x \in (0,\frac{\pi}{4})$.

        $\mathrm{(ii)}$ $w(x,\alpha)$ is strictly concave for $\alpha \in (1,2)$ with respect to $x \in (0,\frac{\pi}{16}]$.
    \end{lemma}
    We leave the proof of Lemma \ref{lem7} in Appendix \ref{A.9.}.
    \begin{proposition}\label{pro5} In qubit systems, for $\alpha \in (1,2),\beta \in (-\infty,0) \cup (0,1),$ there exists some $U_{0} \in \mathcal{U}(\mathcal{H})$ and input state $|\psi_{0} \rangle$ such that
        \begin{equation}\label{eq27}
            M_{\alpha,\beta}(U_{0}|\psi_{0}\rangle)-M_{\alpha,\beta}(|\psi_{0}\rangle) > \mathcal{M}_{\alpha,\beta}(U_{0}).
        \end{equation}
    \end{proposition}
    We leave the proof of Proposition \ref{pro5} in Appendix
    \ref{A.10.}. This result shows that the initial nonstabilizerness in
    the input state can enhance the magic generating power when
    considering $\mathcal{M}_{\alpha,\beta}$ with $\alpha \in
    (1,2),\beta \in (-\infty,0) \cup (0,1)$ in qubit systems.
    \section{Examples}\label{sec6}
\begin{figure}[ht]\centering
        \subfigure[] {\begin{minipage}[Figure1a]{0.42\linewidth}
                \includegraphics[width=1.0\textwidth]{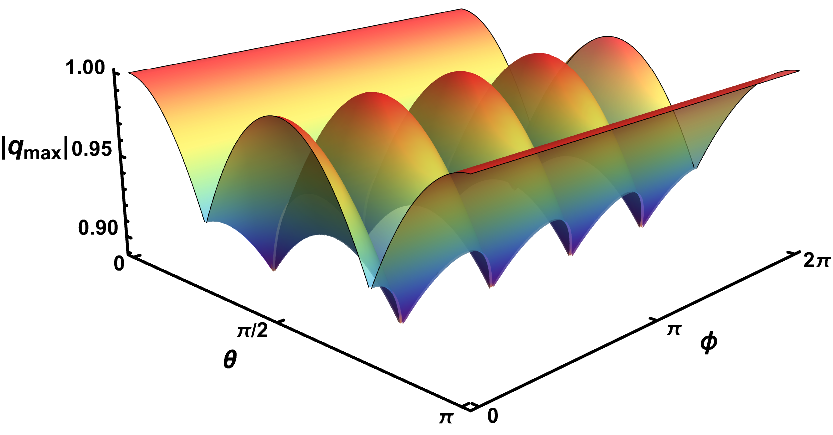}
        \end{minipage}}
        \subfigure[] {\begin{minipage}[Figure1b]{0.55\linewidth}
                \includegraphics[width=1.0\textwidth]{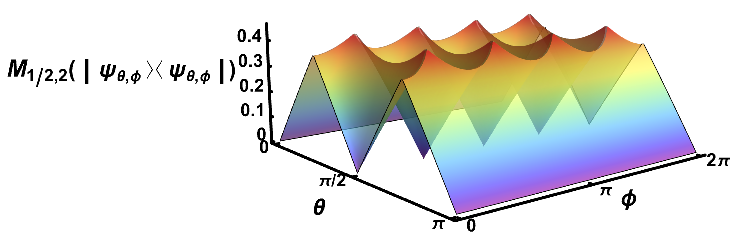}
        \end{minipage}}
        \caption{{The surfaces of $|q_{\max}|$ and $M_{1/2,2}\left( |\psi_{\theta,\phi} \rangle \langle \psi_{\theta,\phi}|\right)$ with the variation of $\theta \in [0,\pi]$ and $\phi \in [0,2\pi)$, respectively.}}
        \label{fig:example1}
    \end{figure}

    In this section, we give three detailed examples to illustrate our results.
    \vskip0.1in
    \noindent {\bf Example 1.}
    Note that in qubit systems,
    $
    \mathcal{PS}_{2}=\left\lbrace |0 \rangle,|1 \rangle ,|+ \rangle ,|- \rangle, |+\mathrm{i} \rangle ,|-\mathrm{i} \rangle \right\rbrace
    $,
    and any pure state $|\psi \rangle$ can be represented as
    $|\psi\rangle=|\psi_{\theta,\phi} \rangle=\mathrm{cos}
    \frac{\theta}{2} |0 \rangle+ \text{e}^{\mathrm{i}\phi} \mathrm{sin}
    \frac{\theta}{2} |1 \rangle$ with $(\theta,\phi) \in [0,\pi] \times
    (0,2\pi]$. Then we have
    \begin{equation}\label{eq28}
        M_{\alpha, \beta}  \left( |\psi_{\theta,\phi} \rangle \langle \psi_{\theta,\phi}|\right)=\frac{1}{(1-\alpha)\beta}\left\lbrace \left[  \left(\frac{1+|q_{\max}|}{2} \right)^{\alpha}+\left(\frac{1-|q_{\max}|}{2} \right) ^{\alpha}    \right] ^{\beta}-1\right\rbrace,
    \end{equation}
    where
    \begin{align}\label{eq29}
        |q_{\max}|
        =&\mathrm{max}\left\{\left| \mathrm{cos} \frac{\theta}{2}\right| ,\left| \mathrm{sin} \frac{\theta}{2}\right| ,\sqrt{\frac{1+\mathrm{sin}\theta \mathrm{cos} \phi}{2}}, \nonumber\right. \\ &\left.\sqrt{\frac{1-\mathrm{sin}\theta \mathrm{cos} \phi}{2}},\sqrt{\frac{1+\mathrm{sin}\theta \sin \phi}{2}},\sqrt{\frac{1-\mathrm{sin}\theta \mathrm{sin} \phi}{2}}\right\}.
    \end{align}
    Thus we obtain
    \begin{equation}\label{eq30}
        M_{1/2, 2}  \left( |\psi_{\theta,\phi} \rangle \langle \psi_{\theta,\phi}|\right)=\left[\left( \sqrt{\frac{1+|q_{\max}|}{2}}+\sqrt{\frac{1-|q_{\max}|}{2}}\right) ^{2}-1\right].
    \end{equation}
    We depict  $|q_{\max}|$ and $M_{1/2, 2}  \left( |\psi_{\theta,\phi} \rangle \langle \psi_{\theta,\phi}|\right)$ in Figure \ref{fig:example1}.

    At the same time, we can show that $M_{\alpha, \beta}  (\rho)$ are bounded in qubit systems with $\alpha \in (0,1) \cup (1,+\infty) ,\beta \in (-\infty,0) \cup (0,+\infty)$.

    We first  introduce a vital class of magic states which are called $T$-type states in qubit systems, i.e.,
    \begin{equation}\label{eq31}
        |T_{j,k} \rangle =\mathrm{cos} \frac{\theta_{j}}{2} |0 \rangle+\text{e}^{\mathrm{i}\mathrm{\phi}_k} \mathrm{sin}\frac{\theta_{j}}{2}|1 \rangle,
    \end{equation}
    where $\theta_{0}=\mathrm{arccos}\left( \frac{1}{\sqrt{3}}\right)  \in (0,\frac{\pi}{2}),\theta_{1}=\pi-\theta_{0}$ and $\phi_{k}=\frac{(2k+1)\pi}{4},j=0,1,k=0,1,2,3$.

    Direct calculations show that in qubit systems, we have
    \begin{equation}\label{eq32}
        \frac{1+\sqrt{\frac{3+\sqrt{3}}{6}}}{2} \leq |q_{\max}| \leq 1,
    \end{equation}
    and the upper bound holds iff $\rho \in \mathcal{S}_{d}$, while the lower bound holds iff $\rho$ is a $T$-type state $|T_{j,k} \rangle$ with $j=0,1,k=0,1,2,3$.

    Using Lemma \ref{lem3} (i), Lemma \ref{lem4} and Eqs. (\ref{eq28}) and (\ref{eq32}), we obtain
    \begin{equation}\label{eq33}
        0 \leq M_{\alpha, \beta}  (|\psi_{\theta,\phi} \rangle \langle \psi_{\theta,\phi}|)
        \leq \frac{1}{(1-\alpha)\beta}\left\lbrace \left[ \left(\frac{1+\sqrt{\frac{3+\sqrt{3}}{6}}}{2}\right)^{\alpha} + \left(\frac{1-\sqrt{\frac{3+\sqrt{3}}{6}}}{2}\right)^{\alpha}\right]^{\beta}-1 \right\rbrace.
    \end{equation}

    Suppose that the optimal decomposition of $M_{\alpha,\beta}\left(\rho\right)$ is $\left\lbrace\left(p_{j},|\psi_{j}\rangle\right) \right\rbrace $. Then we have
    \begin{align}\label{eq34}
        0 \leq M_{\alpha,\beta}\left(\rho\right)
        &=\sum_{j} p_{j} M_{\alpha,\beta}\left(|\psi_{j} \rangle \langle \psi_{j}|\right) \nonumber\\
        & \leq  \frac{1}{(1-\alpha)\beta}\left\lbrace \left[ \left(\frac{1+\sqrt{\frac{3+\sqrt{3}}{6}}}{2}\right)^{\alpha}
        +\left(\frac{1-\sqrt{\frac{3+\sqrt{3}}{6}}}{2}\right)^{\alpha}\right]^{\beta}-1 \right\rbrace.
    \end{align}
    The lower bound in Eq. (\ref{eq33}) saturates iff $M_{\alpha,\beta}\left(|\psi_{j} \rangle \langle \psi_{j}|\right)=0$ for all $j$, iff $|\psi_{j} \rangle \in \mathcal{PS}_{d}$ for all $j$, iff $\rho \in \mathcal{S}_{d}$, since $M_{\alpha,\beta}\left(\rho\right) $ is convex for $\alpha \in (0,1) \cup (1,+\infty),\beta \in (-\infty,0)\cup(0,+\infty)$, the upper bound in Eq. (\ref{eq34}) holds iff $\rho$ is a pure state, thus the upper bound in Eq. (\ref{eq34}) saturates iff $\rho$ is a $T$-type state.
    \vskip0.1in
    \noindent {\bf Example 2.} Consider qutrit systems and note that
    \[
    \mathcal{PS}_{3}=\left\lbrace|0\rangle,|1\rangle,|2\rangle \right\rbrace \cup \left\lbrace \frac{|0\rangle+\omega^{j}|1\rangle+\omega^{k}|2\rangle}{\sqrt{3}}:\omega=\text{e}^{\frac{2\pi}{3}\mathrm{i}}, j,k=0,1,2\right\rbrace.
    \]

    The qutrit $T$ state is defined as
    $
    |T\rangle =\frac{1}{\sqrt{3}}\left(\text{e}^{\frac{2\pi }{9}\mathrm{i}} |0\rangle +|1\rangle+\text{e}^{\frac{-2\pi }{9}\mathrm{i}}|2\rangle \right).
    $
    Then by direct calculations, we have
    \[
    |\langle m|T\rangle|=\frac{1}{\sqrt{3}},\,\,
    |\langle \psi_{j,k}|T\rangle|=|\text{e}^{\frac{2\pi }{9}\mathrm{i}}+\text{e}^{-\frac{2\pi j}{3}\mathrm{i}}+\text{e}^{\left(-\frac{2\pi}{9}-\frac{2\pi k}{3}\right)\mathrm{i}}|,\,\, m,j,k=0,1,2,
    \]
    where $|\psi_{j,k}\rangle=\frac{|0\rangle+\omega^{j}|1\rangle+\omega^{k}|2\rangle}{\sqrt{3}},j,k=0,1,2.$ Thus we obtain
    \begin{align*}
        |c_{\max}|
        &=\mathrm{max}\left\lbrace  |\langle m|T\rangle|,|\langle\psi_{j,k}|T\rangle|,m,j,k=0,1,2\right\rbrace \\
        &=|\langle \psi_{0,0}|T\rangle|=|\langle \psi_{0,2}|T\rangle|=|\langle \psi_{2,2}|T\rangle|\\
        &=\frac{1}{3}\left[ 1+\mathrm{cos}\left(\frac{2\pi}{9} \right) \right].
    \end{align*}

    By Proposition \ref{pro2}, we have
    \begin{equation}\label{eq35}
        M_{\alpha, \beta}  \left(|T\rangle \langle T|\right)=\frac{1}{(1-\alpha)\beta}\left\lbrace \left[  \left(\frac{2}{3}+\frac{\mathrm{cos}\left(\frac{2\pi}{9}\right)}{6} \right)^{\alpha}+\left(\frac{1}{3}-\frac{\mathrm{cos}\left(\frac{2\pi}{9}\right)}{6} \right) ^{\alpha}    \right] ^{\beta}-1\right\rbrace,
    \end{equation}
    and
    \begin{equation}\label{eq36}
        m_{\alpha, \beta}  \left(|T\rangle \langle T|\right)=\frac{1}{(1-\alpha)\beta}\left\lbrace 1-\left[  \left(\frac{2}{3}+\frac{\mathrm{cos}\left(\frac{2\pi}{9}\right)}{6} \right)^{2-\alpha}+\left(\frac{1}{3}-\frac{\mathrm{cos}\left(\frac{2\pi}{9}\right)}{6} \right) ^{2-\alpha}  \right] ^{\beta}\right\rbrace
    \end{equation}
    for $\alpha\in(0,1) \cup (1,2),\beta\in(-\infty,0)\cup(0,+\infty)$.

\begin{figure}[H]\centering
        %\subfigure[]
        {\begin{minipage}[figure2]{0.6\linewidth}
                \includegraphics[width=0.85\textwidth]{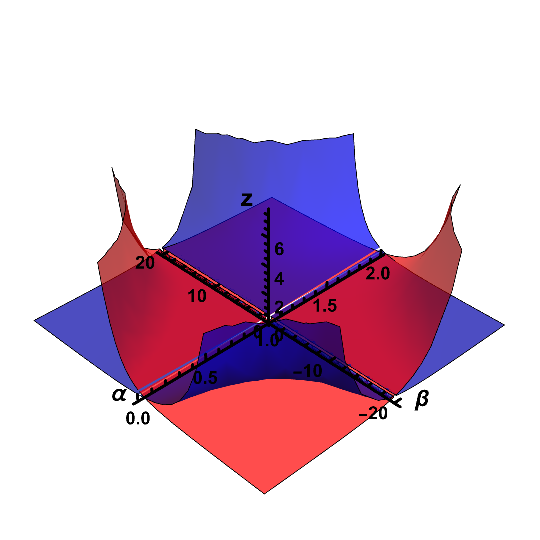}
        \end{minipage}}
        \caption{{The red surface represents $M_{\alpha, \beta}  \left(|T\rangle \langle T|\right)$ and the blue surface represents $m_{\alpha, \beta}  \left(|T\rangle \langle T|\right)$ with $\alpha\in (0,1) \cup (1,2)$ and $\beta \in (-20,0) \cup (0,20)$. }}
        \label{fig:example2}
    \end{figure}

\begin{figure}[H]\centering
        {\begin{minipage}[figure3]{0.6\linewidth}
                \includegraphics[width=0.9\textwidth]{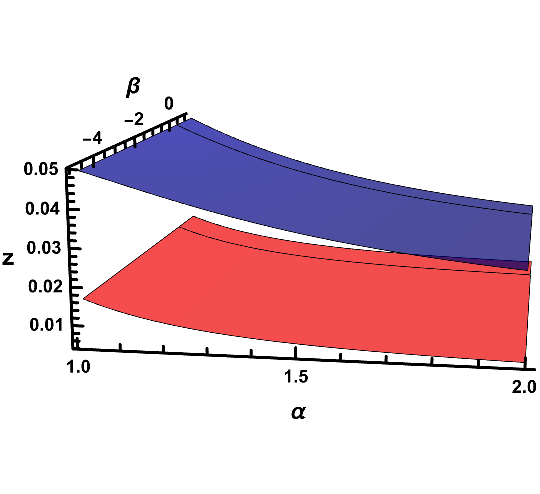}
        \end{minipage}}
        \caption{{The blue surface represents $M_{\alpha,\beta}(T^{1/4}|\psi_{0}\rangle)-M_{\alpha,\beta}(|\psi_{0}\rangle)$ and the red surface represents $\mathcal{M}_{\alpha,\beta}(T^{1/4})$ with $\alpha\in (1,2)$ and $\beta \in (-5,0) \cup (0,1)$. }}
        \label{fig:example3}
    \end{figure}

    We depict $M_{\alpha, \beta}  \left(|T\rangle \langle T|\right)$ and $m_{\alpha, \beta}  \left(|T\rangle \langle T|\right)$ in Eqs. (\ref{eq35}) and (\ref{eq36}) with $\alpha \in (0,1) \cup (1,2)$ and $\beta  \in (-20,0 )\cup (0,20)$ in Figure \ref{fig:example2}. It can be seen that $M_{\alpha,\beta}\left(|T\rangle \langle T|\right)$ and $m_{\alpha,\beta}\left(|T\rangle \langle T|\right)$ are highly symmetric in this parameter range.
    \vskip0.1in
    \noindent {\bf Example 3.} According to the proof of Proposition \ref{pro5}, we have
    \begin{align*}
        M_{\alpha,\beta}(T^{1/4}|\psi_{0}\rangle)-M_{\alpha,\beta}(|\psi_{0}\rangle)
        =&\frac{1}{(1-\alpha)\beta}\left\lbrace \left[\cos^{2\alpha}\left(\frac{3\pi}{64}\right)+\sin^{2\alpha}\left(\frac{3\pi}{64}\right)\right]^{\beta} \right.\\
        &- \left. \left[\cos^{2\alpha}\left(\frac{\pi}{32}\right)+\sin^{2\alpha}\left(\frac{\pi}{32}\right)\right]^{\beta}
        \right\rbrace ,
    \end{align*}
    and
    \[
    \mathcal{M}_{\alpha,\beta}(T^{1/4})=\frac{1}{(1-\alpha)\beta}\left\lbrace \left[ \cos^{2\alpha}\left(\frac{\pi}{64}\right)+\sin^{2\alpha}\left(\frac{\pi}{64}\right)\right]^{\beta}-1 \right\rbrace
    \]
    for $\alpha\in (1,2)$ and $\beta \in (-5,0) \cup (0,1)$, where $T$ is a qubit $T$-gate, and $ |\psi_{0} \rangle =\frac{1}{\sqrt{2}}\left(|0 \rangle + \text{e}^{\frac{\pi }{8}\mathrm{i}}|1\rangle\right)$ in qubit systems.
    We depict $M_{\alpha,\beta}(T^{1/4}|\psi_{0}\rangle)-M_{\alpha,\beta}(|\psi_{0}\rangle)$ and $\mathcal{M}_{\alpha,\beta}(T^{1/4})$ with $\alpha\in (1,2)$ and $\beta \in (-5,0) \cup (0,1)$ in Figure \ref{fig:example3}. It shows that $M_{\alpha,\beta}(T^{1/4}|\psi_{0}\rangle)-M_{\alpha,\beta}(|\psi_{0}\rangle)$ is always larger than $\mathcal{M}_{\alpha,\beta}(T^{1/4})$ in this parameter range.

    \section{Summary}\label{sec7}
    We have proposed two pure-state stabilizer monotones which can be
    calculated on low-dimensional systems for pure states. We have
    showed that the quantifier $M_{\alpha,\beta}(\rho)$ is bounded in
    qubit systems with $\alpha \in (0,1) \cup (1,+\infty)$ and $\beta
    \in (-\infty,0) \cup (0,+\infty)$, and the lower bound holds if and
    only if $\rho \in \mathcal{S}_{d}$, while the upper bound holds if
    and only if $\rho$ is a $T$-type state. We have also compared our
    magic quantifiers with other quantifiers, showing that
    $M_{\alpha,\beta}$ is controlled by some function of the robustness
    of magic $R$ for pure states with $\alpha \in (0,1) \cup
    (1,+\infty), \beta \in (1,+\infty)$. As an application, we have
    proved that the initial nonstabilizerness in the input state can
    enhance the magic generating power in qubit systems when considering
    $\mathcal{M}_{\alpha,\beta}$ for $\alpha \in (1,2),\beta \in
    (-\infty,0) \cup (0,1)$. Our results may shed some new light on the
    study of quantification of the magic at the level of quantum states,
    and may offer new perspective in the research of magic resource
    theory.
    \vskip0.1in

    \noindent

    \subsubsection*{Acknowledgements}
    This work was supported by National Natural Science Foundation of
    China (Grant Nos. 12561084,12161056) and Natural Science Foundation of Jiangxi
    Province of China (Grant No. 20232ACB211003).

    \subsubsection*{Author Contributions}
    Linmao Wang wrote the main manuscript text and Zhaoqi Wu
    supervised and revised the manuscript. All authors reviewed the
    manuscript.

    \subsubsection*{Data Availability}
    No datasets were generated or analysed during the current study.

    \subsubsection*{Competing interests}
    The authors declare no competing interests.

    \appendix
    \section*{Appendix}
    \refstepcounter{section}
    Here we give the proof of Proposition \ref{pro1}-\ref{pro5}, Theorem \ref{thm1}-\ref{thm5} and Lemma \ref{lem7}.
    \subsection{Proof of Proposition 1} \label{A.1.}

    Applying the spectral decomposition, we obtain
    \[
    \frac{|\psi \rangle \langle \psi|+|\phi \rangle \langle \phi|}{2}=\lambda_{1}|\xi_{1} \rangle \langle \xi_{1}|+\lambda_{2}|\xi_{2} \rangle \langle \xi_{2}|,
    \]
    where
    \[
    \lambda_1=\frac{1+|\langle \psi|\phi \rangle|}{2},\,\,\lambda_2=\frac{1-|\langle \psi|\phi \rangle|}{2},
    \]
    \[
    |\xi_1\rangle = \frac{|\psi\rangle + \text{e}^{\mathrm{i}\theta}|\phi\rangle}{\sqrt{2(1+|s|)}}, \,\,|\xi_2\rangle = \frac{|\psi\rangle - \text{e}^{\mathrm{i}\theta}|\phi\rangle}{\sqrt{2(1-|s|)}},
    \]
    \[
    s=\langle \phi | \psi \rangle, \,\,s=|s|\text{e}^{\mathrm{i}\theta}, \theta \in \mathbb{R}.
    \]
    Hence we have
    \begin{align}
        &J_{\alpha, \beta}\left( |\psi \rangle \langle \psi|,|\phi \rangle \langle \phi|\right) \notag\\
        =& S_{\alpha, \beta}\left(\frac{|\psi \rangle \langle \psi|+|\phi \rangle \langle \phi|}{2}\right)-\frac{1}{2}S_{\alpha, \beta}\left( |\psi \rangle \langle \psi|\right) -\frac{1}{2}S_{\alpha, \beta}\left( |\phi \rangle \langle \phi|\right) \notag\\
        =&S_{\alpha, \beta}\left(\frac{|\psi \rangle \langle \psi|+|\phi \rangle \langle \phi|}{2}\right) \notag\\
        =&H_{\alpha, \beta}\left\lbrace \left( \frac{1+|\langle \psi|\phi \rangle|}{2},\frac{1-|\langle \psi|\phi \rangle|}{2}\right) \right\rbrace \notag\\
        =&\frac{1}{(1-\alpha)\beta}\left[ \left(\lambda_{1}^{\alpha}+\lambda_{2}^{\alpha}\right)^{\beta}-1\right],
        \tag{A1} \label{eq:A1}
    \end{align}
    where
    \begin{equation}\tag{A2} \label{eq:A2}
        H_{\alpha, \beta}(\left\lbrace p_{i} \right\rbrace) =\frac{1}{(1-\alpha)\beta}\left[ \left( \sum_{j}^{}p_i^{\alpha}\right)^{\beta} -1\right]
    \end{equation}
    is the unified $(\alpha,\beta)$-entropy\cite{Rathie}.

    Noting that
    \[
    |\langle \psi | \xi_{i} \rangle|^{2}=|\langle \phi | \xi_{i} \rangle|^{2}=\lambda_{i},i=1,2,
    \]
    we have
    \begin{align}
        &D_{\alpha, \beta} \left(|\psi \rangle \langle \psi| \Big\|\frac{|\psi \rangle \langle \psi|+|\phi \rangle \langle \phi|}{2}\right) \notag \\
        =&\frac{1}{(1-\alpha)\beta}\left\lbrace 1-
        \left[  \mathrm{Tr} \left( (|\psi \rangle \langle \psi|)^{\alpha}\left(\frac{|\psi \rangle \langle \psi|+|\phi \rangle \langle \phi|}{2}\right)^{1-\alpha} \right) \right] ^{\beta} \right\rbrace  \notag\\
        =&\frac{1}{(1-\alpha)\beta}\left[ 1- \left(\lambda_{1}^{1-\alpha}|\langle \psi | \xi_{1} \rangle|^{2}+\lambda_{2}^{1-\alpha}|\langle \psi | \xi_{2} \rangle|^{2}\right)^{\beta}\right] \notag \\
        =&\frac{1}{(1-\alpha)\beta} \left[ 1- \left(\lambda_{1}^{2-\alpha}+\lambda_{2}^{2-\alpha}\right)^{\beta}\right],
        \tag{A3} \label{eq:A3}
    \end{align}
    and
    \begin{align}
        &D_{\alpha, \beta} \left(|\phi \rangle \langle \phi|\Big\|\frac{|\psi \rangle \langle \psi|+|\phi \rangle \langle \phi|}{2}\right)\notag \\
        =&\frac{1}{(1-\alpha)\beta}\left\lbrace 1-\left[  \mathrm{Tr}\left(  (|\phi \rangle \langle \phi|)^{\alpha}\left(\frac{|\psi \rangle \langle \psi|+|\phi \rangle \langle \phi|}{2}\right)^{1-\alpha} \right) \right] ^{\beta} \right\rbrace  \notag\\
        =&\frac{1}{(1-\alpha)\beta}\left[ 1- \left(\lambda_{1}^{1-\alpha}|\langle \phi | \xi_{1} \rangle|^{2}+\lambda_{2}^{1-\alpha}|\langle \phi | \xi_{2} \rangle|^{2}\right)^{\beta}\right] \notag\\
        =&\frac{1}{(1-\alpha)\beta} \left[ 1- \left(\lambda_{1}^{2-\alpha}+\lambda_{2}^{2-\alpha}\right)^{\beta}\right]
        \tag{A4} \label{eq:A4}.
    \end{align}
    Combining Eqs. (\ref{eq:A3}) and (\ref{eq:A4}), we obtain
    \begin{equation}\tag{A5} \label{eq:A5}
        J_{\alpha, \beta}^{\prime} \left( |\psi \rangle \langle \psi|,|\phi \rangle \langle \phi|\right)=\frac{1}{(1-\alpha)\beta} \left[ 1- \left(\lambda_{1}^{2-\alpha}+\lambda_{2}^{2-\alpha}\right)^{\beta}\right],
    \end{equation}
    and thus we get
    \[
    J_{\alpha, \beta}\left( |\psi \rangle \langle \psi|,|\phi \rangle \langle \phi|\right)=J_{2-\alpha, \beta}^{\prime} \left( |\psi \rangle \langle \psi|,|\phi \rangle \langle \phi|\right),
    \]
    for all $\alpha \in (0,1)\cup (1,2)$ and $\beta \in (-\infty,0)\cup (0,+\infty)$ by comparing Eqs. (\ref{eq:A1}) and (\ref{eq:A5}). Thus we complete the proof. $\hfill\qedsymbol$
    \subsection{Proof of Theorem 1} \label{A.2.}

    (i)-(iii) can be directly proved by applying Lemma \ref{lem1}, and (iv) is obvious from the definition of $J_{\alpha,\beta}(\rho,\sigma)$, so we only prove item (v).

    For $\alpha \in (1,+\infty) , \beta \in [1,+\infty)$, we have
    \begin{align*}
        &|J_{\alpha,\beta}(\rho_1,\sigma)-J_{\alpha,\beta}(\rho_2,\sigma)|\\
        =&\left|S_{\alpha,\beta}\left(\frac{\rho_{1}+\sigma}{2}\right)- S_{\alpha,\beta}\left(\frac{\rho_{2}+\sigma}{2}\right) +\frac{1}{2}S_{\alpha,\beta}(\rho_{2})-\frac{1}{2}S_{\alpha,\beta}(\rho_{1})\right| \\
        \leq&\left|S_{\alpha,\beta}\left(\frac{\rho_{1}+\sigma}{2}\right)- S_{\alpha,\beta}\left(\frac{\rho_{2}+\sigma}{2}\right)\right| +\frac{1}{2} \left| S_{\alpha,\beta}(\rho_{2})-S_{\alpha,\beta}(\rho_{1})\right| \\
        \leq&\frac{\alpha}{\alpha-1} \left\| \frac{\rho_{1}-\rho_{2}}{2}\right\| _{1}+ \frac{\alpha}{2(\alpha-1)} \left\| \rho_{2}-\rho_{1}\right\| _{1}\\
        =&\frac{\alpha}{\alpha-1} \left\| \rho_{1}-\rho_{2}\right\| _{1},
    \end{align*}
    where the first equality follows from Eq. (\ref{eq1}), the second inequality comes from Lemma \ref{lem1} (iv), and the second equality follows from the linearity of the trace distance.

    In a similar manner, we can prove that $|J_{\alpha,\beta}(\rho,\sigma_{1})-J_{\alpha,\beta}(\rho,\sigma_{2})| \leq \frac{\alpha}{\alpha-1}\|\sigma_{1}-\sigma_{2}\|_{1}$ for $\alpha \in (1,+\infty) , \beta \in [1,+\infty)$. $\hfill\qedsymbol$
    \subsection{Proof of Theorem 2}\label{A.3.}

    We can prove (iii-v) by Lemma \ref{lem2} (iii-v) respectively and (vi) can be obtained by definition, so we only prove (i) and (ii).

    For item (i), by Lemma \ref{lem2} (i), we have $J_{\alpha, \beta}^{\prime}(\rho, \sigma) \geq 0$ for $\alpha \in (0,1)\cup (1,+\infty),\beta \in (-\infty,0) \cup (0,+\infty)$ and the inequality saturates if and only if
    \[
    D_{\alpha, \beta} \left(\rho\Big\|\frac{\rho+\sigma}{2}\right)=0=D_{\alpha, \beta} \left(\sigma\Big\|\frac{\rho+\sigma}{2}\right),
    \]
    which is equivalent to $\rho=\sigma$.

    For item (ii), we have
    \begin{align*}
        &J_{\alpha, \beta}^{\prime}(\rho \otimes \tau, \sigma \otimes \tau) \\
        =&\frac{1}{2}\left[D_{\alpha, \beta} \left(\rho \otimes \tau \Big\|\frac{\rho+\sigma}{2} \otimes \tau\right)+D_{\alpha, \beta} \left(\sigma \otimes \tau \Big\|\frac{\rho+\sigma}{2} \otimes \tau \right)\right]\\
        =&\frac{1}{2} \left[D_{\alpha, \beta} \left(\rho\Big\|\frac{\rho+\sigma}{2}\right)+(\alpha-1)\beta D_{\alpha, \beta} \left(\rho\Big\|\frac{\rho+\sigma}{2}\right)D_{\alpha, \beta} \left(\tau\|\tau\right)  \right]\\
        &+\frac{1}{2} \left[D_{\alpha, \beta} \left(\sigma\Big\|\frac{\rho+\sigma}{2}\right)+(\alpha-1)\beta D_{\alpha, \beta} \left(\sigma\Big\|\frac{\rho+\sigma}{2}\right)D_{\alpha, \beta} \left(\tau\|\tau \right)  \right]\\
        =&\frac{1}{2}\left[D_{\alpha, \beta} \left(\rho\Big\|\frac{\rho+\sigma}{2}\right)+D_{\alpha, \beta} \left(\sigma\Big\|\frac{\rho+\sigma}{2}\right)\right]\\
        =&J_{\alpha, \beta}^{\prime}(\rho, \sigma),
    \end{align*}
    with $\alpha \in (0,1)\cup (1,+\infty),\beta \in (-\infty,0) \cup
    (0,+\infty)$, where the first and last equality comes from Eq.
    (\ref{eq2}), the second equality follows from Lemma \ref{lem2} (ii),
    and the third holds due to $D_{\alpha,\beta}(\tau\|\tau)=0$.
    $\hfill\qedsymbol$
    \subsection{Proof of Theorem 3}\label{A.4.}

    We only prove item (iv) and (vi), the other properties can be easily derived by imitating the proof of proposition 1 in \cite{Tian1}.

    For item (iv), suppose that $\alpha \in (1,2), \beta \in (-\infty,1]$. Then we have
    \begin{align*}
        &M_{\alpha, \beta} \left(\mathcal{E}\left( |\psi \rangle \langle \psi|\right) \right)\\
        =&\min_{|\phi \rangle \in \mathcal{PS}_{d}} J_{\alpha,\beta} \left( \mathcal{E}\left( |\psi \rangle \langle \psi|\right) , |\phi \rangle \langle \phi|\right) \\
        \leq&\min_{|\phi \rangle \in \mathcal{PS}_{d}} J_{\alpha,\beta} \left( \mathcal{E}\left( |\psi \rangle \langle \psi|\right) ,\mathcal{E}(|\phi \rangle \langle \phi|)\right) \\
        \leq& \min_{|\phi \rangle \in \mathcal{PS}_{d}}J_{\alpha, \beta} \left( |\psi \rangle \langle \psi|,|\phi \rangle \langle \phi|\right) \\
        =&M_{\alpha, \beta}  \left(|\psi \rangle \langle \psi|\right),
    \end{align*}
    where the first equality holds by using Eq. (\ref{eq5}) and the fact that any stabilizer operation $\mathcal{E}$ maps pure states to pure states, the first inequality is true since stabilizer operation preserves stabilizer states and the second inequality follows from Remark \ref{rem2}.

    For item (vi), suppose that $\alpha \in (1,+\infty) , \beta \in [1,+\infty)$. Then we have
    \begin{align*}
        &M_{\alpha,\beta}(|\psi_{1} \rangle \langle \psi_{1}|)-M_{\alpha,\beta}(|\psi_{2} \rangle \langle \psi_{2}|)\\
        =&\min_{|\phi \rangle \in \mathcal{PS}_{d}} J_{\alpha,\beta} \left(  |\psi_{1} \rangle \langle \psi_{1}| , |\phi \rangle \langle \phi|\right) - J_{\alpha,\beta} \left(  |\psi_{2} \rangle \langle \psi_{2}| , |\phi_{2} \rangle \langle \phi_{2}|\right) \\
        \leq&  J_{\alpha,\beta} \left(  |\psi_{1} \rangle \langle \psi_{1}| , |\phi_{2} \rangle \langle \phi_{2}|\right) - J_{\alpha,\beta} \left(  |\psi_{2} \rangle \langle \psi_{2}| , |\phi_{2} \rangle \langle \phi_{2}|\right)  \\
        \leq&  |J_{\alpha,\beta} \left(  |\psi_{1} \rangle \langle \psi_{1}| , |\phi_{2} \rangle \langle \phi_{2}|\right) - J_{\alpha,\beta} \left(  |\psi_{2} \rangle \langle \psi_{2}| , |\phi_{2} \rangle \langle \phi_{2}|\right) | \\
        \leq& \frac{\alpha}{\alpha-1} \| |\psi_{1} \rangle \langle \psi_{1}|-|\psi_{2} \rangle \langle \psi_{2}| \|_{1},
    \end{align*}
    where the first equality holds because we assume that $|\phi_{2}\rangle$ reaches the minimum in the definition of $M_{\alpha,\beta}(|\psi_{2} \rangle \langle \psi_{2}|)$, the last inequality comes from Theorem \ref{thm1} (v). Similarly,
    we can obtain that $M_{\alpha,\beta}(|\psi_{2} \rangle \langle \psi_{2}|)-M_{\alpha,\beta}(|\psi_{1} \rangle \langle \psi_{1}|) \leq \frac{\alpha}{\alpha-1} \| |\psi_{1} \rangle \langle \psi_{1}|-|\psi_{2} \rangle \langle \psi_{2}| \|_{1} $, and thus we have
    \[
    |M_{\alpha,\beta}(|\psi_{1} \rangle \langle \psi_{1}|)-M_{\alpha,\beta}(|\psi_{2} \rangle \langle \psi_{2}|)| \leq \frac{\alpha}{\alpha-1} \| |\psi_{1} \rangle \langle \psi_{1}|-|\psi_{2} \rangle \langle \psi_{2}| \|_{1},
    \]
    which completes the proof. $\hfill\qedsymbol$
    \subsection{Proof of Proposition 2} \label{A.5.}

    For $\alpha \in (0,1),\beta \in (-\infty,0) \cup (0,+\infty)$, we have
    \begin{align*}
        &M_{\alpha, \beta}  \left( |\psi \rangle \langle \psi|\right)\\
        =&\min_{|\phi \rangle \in \mathcal{PS}_{d}}\left\lbrace \frac{1}{(1-\alpha)\beta}\left[ \left(\left( \frac{1+| \langle \phi  |\psi \rangle|}{2}\right)^{\alpha}+\left( \frac{1-| \langle \phi  |\psi \rangle|}{2}\right)^{\alpha}\right)^{\beta}-1\right]\right\rbrace \\
        =&\frac{1}{(1-\alpha)\beta}\left\lbrace \left[ \min_{|\phi \rangle \in \mathcal{PS}_{d}} \left(\left(\frac{1+| \langle \phi  |\psi \rangle|}{2} \right)^{\alpha}+\left(\frac{1-| \langle \phi  |\psi \rangle|}{2} \right) ^{\alpha}  \right)  \right] ^{\beta}-1\right\rbrace \\
        =&\frac{1}{(1-\alpha)\beta}\left\lbrace \left[  \left(\frac{1+|c_{\psi}|}{2} \right)^{\alpha}+\left(\frac{1-|c_{\psi}|}{2} \right) ^{\alpha}    \right] ^{\beta}-1\right\rbrace,
    \end{align*}
    the first equality follows from the definition of $M_{\alpha, \beta}  \left( |\psi \rangle \langle \psi|\right)$, the second equality comes from Lemma \ref{lem3} $(\mathrm{i})$ and the third follows from Lemma \ref{lem4}.

    For $\alpha \in (1,+\infty),\beta \in (-\infty,0) \cup (0,+\infty)$, we only need to replace min with max in the second equality to obtain the same conclusion. Thus we complete the proof. $\hfill\qedsymbol$
    \subsection{Proof of Proposition 3} \label{A.6.}

    We only consider the case in which $\alpha \in (1,+\infty)$, while the proof is similar when $\alpha \in (0,1)$. Denote
    \[
    h(\lambda,\alpha,\beta)=\frac{1}{(1-\alpha)\beta}\left\{\left[\lambda^{\alpha}+\left(1-\lambda\right)^{\alpha}\right]^{\beta}-1\right\}.
    \]

    For $\alpha \in (1,+\infty)$, by applying Lemma \ref{lem6}, we have
    \[
    \left[ (2\lambda-1)^{2}+h(\lambda,\alpha,\beta)\right] \leq \lim\limits_{\beta \rightarrow 1^{+}} \left[ (2\lambda-1)^{2}+h(\lambda,\alpha,\beta)\right].
    \]
    then by Lemma \ref{lem5} and proposition 3 in \cite{Tian1}, we obtain
    \[
    \lim\limits_{\beta \rightarrow 1^{+}} \left[(2\lambda-1)^{2}+h(\lambda,\alpha,\beta)\right] \leq \lim\limits_{\alpha \rightarrow 1^{+}} \lim\limits_{\beta \rightarrow 1^{+}} \left[(2\lambda-1)^{2}+h(\lambda,\alpha,\beta)\right] \leq t_{0},
    \]
    which yields
    \[
    \left[ (2\lambda-1)^{2}+h(\lambda,\alpha,\beta)\right]  \leq t_{0}.
    \]

    From the definition of $M_{\alpha,\beta}\left(|\psi\rangle \langle \psi |\right)$ and $F\left(|\psi \rangle \langle \psi |\right)$,  we obtain
    \[
    M_{\alpha,\beta}\left(|\psi\rangle \langle \psi |\right) + F\left(|\psi \rangle \langle \psi |\right) \leq t_{0},
    \]
    thus
    \[
    M_{\alpha,\beta}\left(|\psi\rangle \langle \psi |\right) +\frac{1}{1+R\left(|\psi \rangle \langle \psi |\right)} \leq M_{\alpha,\beta}\left(|\psi\rangle \langle \psi |\right) + F\left(|\psi \rangle \langle \psi |\right)   \leq
    t_{0},
    \]
    which completes the proof. $\hfill\qedsymbol$
    \subsection{Proof of Theorem 5} \label{A.7.}

    Items (i) and (ii) are obvious from the definition, so we only prove item $\mathrm{(iii)}$.

    For $\alpha \in (1,2), \beta \in (-\infty,0)\cup (0,1]$, we have
    \begin{align*}
        &\mathcal{M}_{\alpha,\beta}(U_{1} \otimes U_{2})\\
        =&\max_{ \rho \in \mathcal{S}_{d^{2}}}  M_{\alpha,\beta}\left[(U_{1}\otimes U_{2})\rho (U_{1}^\dagger \otimes U_{2}^\dagger)\right]  \\
        \geq& \max_{ \rho_{1},\rho_{2} \in \mathcal{S}_{d}}  M_{\alpha,\beta}\left[(U_{1}\otimes U_{2})\left( \rho_{1} \otimes \rho_{2}\right) (U_{1}^\dagger \otimes U_{2}^\dagger)\right]  \\
        =& \max_{ \rho_{1}, \rho_{2} \in \mathcal{S}_{d}}  M_{\alpha,\beta} \left[ \left(U_{1} \rho_{1} U_{1} ^\dagger \right) \otimes \left( U_{2} \rho_{2} U_{2}^\dagger \right)  \right]   \\
        \geq& \max_{ \rho_{1} \in \mathcal{S}_{d}}   M_{\alpha,\beta}\left(U_{1} \rho_{1} U_{1} ^\dagger\right)  \\
        =& \mathcal{M}_{\alpha,\beta}\left(U_{1}\right),
    \end{align*}
    where the first equality holds due to Eq. (\ref{eq23}), the first inequality is true since $\mathcal{S}_{d} \otimes \mathcal{S}_{d} \subseteq \mathcal{S}_{d^{2}}$, and the second inequality follows from Theorem \ref{thm3} (v).
    \subsection{Proof of Proposition 4} \label{A.8.}

    For $\alpha \in (0,1) , \beta \in (-\infty ,0)\cup(0,+\infty)$, we have
    \begin{align*}
        &\mathcal{M}_{\alpha,\beta}(U)\\
        =&\max_{ |\phi \rangle \in \mathcal{PS}_{d}}M_{\alpha,\beta}(U |\phi \rangle \langle \phi|U^{\dagger})\\
        =&\max_{ |\phi \rangle \in \mathcal{PS}_{d}}\min_{|\psi \rangle \in \mathcal{PS}_{d}}\left\lbrace \frac{1}{(1-\alpha)\beta}\left[ \left(\left( \frac{1+| \langle\psi|U|\phi\rangle|}{2}\right)^{\alpha}+\left( \frac{1-| \langle \psi |U |\phi\rangle|}{2}\right)^{\alpha}\right)^{\beta}-1\right]\right\rbrace \\
        =&\frac{1}{(1-\alpha)\beta}\left\lbrace \left[  \left(\frac{1+|C_{U}|}{2} \right)^{\alpha}+\left(\frac{1-|C_{U}|}{2} \right) ^{\alpha}    \right]
        ^{\beta}-1\right\rbrace,
    \end{align*}
    where the first equality follows from the fact that
    $\mathcal{S}_{d}$ is a convex set, the second equation comes from
    the definition, and the last follows from  Lemma \ref{lem3}
    $\mathrm{(i)}$ and Lemma \ref{lem4}.

    It can be easily verified that the conclusion also holds for $\alpha
    \in (1,+\infty), \beta \in (-\infty ,0)\cup(0,+\infty)$. Hence we
    complete the proof. $\hfill\qedsymbol$
    \subsection{Proof of Lemma 7} \label{A.9.}

    (i) For $\alpha \in (1,2),x \in (0,\frac{\pi}{4})$, it follows
    that
    \[
    \frac{\partial w(x,\alpha)}{\partial x}=2 \alpha \cdot  \mathrm{sin}x \cdot \mathrm{cos}x \left(\mathrm{sin}^{2\alpha-2}x-\mathrm{cos}^{2\alpha-2}x\right) <0.
    \]

    (ii)  For $\alpha \in (1,2),x \in (0,\frac{\pi}{16}]$, we obtain
    \[
    \frac{\partial ^{2} w(x,\alpha)}{\partial x^{2}}=2\alpha\Big\{(2\alpha-1)[\sin^{2\alpha-2}x\cos^2x+\cos^{2\alpha-2}x\sin^2x] - (\sin^{2\alpha}x+\cos^{2\alpha}x)\Big\}.
    \]

    Let $t=\tan^{2}x \in \left(0,t_{\max}\right]$, where $t_{\max}=\tan^{2}\frac{\pi}{16}$. Then we have
    \[
    \frac{\partial ^{2} w(x,\alpha)}{\partial x^{2}}=G(t,\alpha)=(2\alpha-1)\left(t^{\alpha-1}+t\right)-\left(t^{\alpha}+1\right).
    \]
    We next show that $G(t,\alpha) <0$ for all $t \in (0,t_{\max}]$ and $\alpha \in (1,2)$.

    By taking the partial derivative of $G(t,\alpha)$ with respect to $t$, we obtain
    \begin{align*}
        \frac{\partial G(t,\alpha)}{\partial t}
        &=(2\alpha-1)(\alpha-1)t^{\alpha-2}+2\alpha-1-\alpha t^{\alpha-1}\\
        & \geq (2\alpha-1)(\alpha-1)+2\alpha-1-\alpha t^{\alpha-1}\\
        &=(2\alpha-1)(\alpha-1)+\left[ \alpha(2-t^{\alpha-1})-1\right]>0.
    \end{align*}
    Then for all $\alpha \in (1,2)$, $G(t,\alpha)_{\max}=G(t_{\max},\alpha)=(2\alpha-1)\left(t^{\alpha-1}_{\max}+t_{\max}\right)-\left(t^{\alpha}_{\max}+1\right).$ So we only need to prove $G(t_{\max},\alpha)<0$ for all $\alpha \in (1,2)$.

    In fact, by taking the first and second derivative of $G(t_{\max},\alpha)$, we have
    \[
    G^{\prime}(t_{\max},\alpha)=2\left( t_{\max}^{\alpha-1}+t_{\max}\right)+\ln t_{\max}\left[ (2\alpha-1)t_{\max}^{\alpha-1}-t_{\max}^{\alpha}\right],
    \]
    \[
    G^{\prime \prime}(t_{\max},\alpha)=t_{\max}^{\alpha-1} \ln t_{\max} \left[(2\alpha-1-t_{\max} ) \ln t_{\max}+4\right].
    \]
    Noting that
    \[
    G^{\prime \prime}(t_{\max},\alpha)>0 \hspace{0.5em} \text{for} \hspace{0.5em} \alpha \in \left(\frac{1+t_{\max}}{2}-\frac{2}{\ln t_{\max}},2\right),
    \]
    and
    \[
    G^{\prime \prime}(t_{\max},\alpha)<0 \hspace{0.5em} \text{for} \hspace{0.5em} \alpha \in \left(1,\frac{1+t_{\max}}{2}-\frac{2}{\ln t_{\max}}\right),
    \]
    we have
    \[
    G^{\prime}(t_{\max},\alpha)_{\max}=\max \left\{G^{\prime}(t_{\max},1),G^{\prime}(t_{\max},2)\right\}=G^{\prime}(t_{\max},2) \approx -0.2205 <0,
    \]
    and so
    \[
    G(t_{\max},\alpha)<\lim\limits_{\alpha \rightarrow 0^{+}}G(t_{\max},\alpha)=0
    \]
    for all $\alpha \in (1,2)$. Thus we complete the proof.
    $\hfill\qedsymbol$

    \subsection{Proof of Proposition 5} \label{A.10.}

    Consider the unitary gate
    \[
    U_{0}=T^{1/4}=\begin{bmatrix}
        1 & 0 \\
        0 & \text{e}^{\frac{\pi}{16}\mathrm{i}}
    \end{bmatrix},
    \]
    and the input state $ |\psi_{0} \rangle =\frac{1}{\sqrt{2}}\left(|0 \rangle + \text{e}^{\frac{\pi }{8}\mathrm{i}}|1\rangle\right).$

    By Proposition \ref{pro1} and Proposition \ref{pro4}, we obtain
    \[
    \mathcal{M}_{\alpha,\beta}(T^{1/4})=\frac{1}{(1-\alpha)\beta}\left\lbrace \left[ \cos^{2\alpha}\left(\frac{\pi}{64}\right)+\sin^{2\alpha}\left(\frac{\pi}{64}\right)\right]^{\beta}-1 \right\rbrace,
    \]
    \[
    M_{\alpha,\beta}(T^{1/4}|\psi_{0}\rangle)=\frac{1}{(1-\alpha)\beta}\left\lbrace \left[ \cos^{2\alpha}\left(\frac{3\pi}{64}\right)+\sin^{2\alpha}\left(\frac{3\pi}{64}\right)\right]^{\beta}-1 \right\rbrace,
    \]
    and
    \[
    M_{\alpha,\beta}(|\psi_{0}\rangle)=\frac{1}{(1-\alpha)\beta}\left\lbrace \left[ \cos^{2\alpha}\left(\frac{\pi}{32}\right)+\sin^{2\alpha}\left(\frac{\pi}{32}\right)\right]^{\beta}-1 \right\rbrace.
    \]

    Denote \[N(x,\alpha,\beta)=\frac{1}{(1-\alpha)\beta}\left[\left(\mathrm{cos}^{2\alpha}x+\mathrm{sin}^{2\alpha}x\right)^{\beta}-1 \right]=g\left[w(x,\alpha),\alpha,\beta\right],
    \]
    and
    \[
    K(x,\alpha,\beta)=N\left(x+\frac{\pi}{32},\alpha,\beta\right)-N(x,\alpha,\beta)-N\left(\frac{\pi}{32},\alpha,\beta\right),
    \]
    where $g(x,\alpha,\beta)$ and $w(x,\alpha)$ are defined in Lemma \ref{lem3} and Lemma \ref{lem7}, respectively.

    Thus we have
    \[
    M_{\alpha,\beta}(T^{1/4}|\psi_{0}\rangle)-M_{\alpha,\beta}(|\psi_{0}\rangle)-\mathcal{M}_{\alpha,\beta}(T^{1/4})=K\left(\frac{\pi}{64},\alpha,\beta\right),
    \]
    Now we prove that $K(\frac{\pi}{64},\alpha,\beta)>0$ for all $\alpha \in (1,2)$ and $\beta \in (-\infty,0) \cup (0,1).$

    Taking the second-order partial derivative of $N(x,\alpha,\beta)$ with respect to $x$ , we obtain
    \[
    \frac{\partial^{2}N(x,\alpha,\beta)}{\partial x^{2}}=\frac{\partial^{2}g[w(x,\alpha),\alpha,\beta]}{\partial[w(x,\alpha)]^{2}}\left[\frac{\partial w(x,\alpha)}{\partial x}\right]^{2}+\frac{\partial^{2} w(x,\alpha)}{\partial x^{2}} \frac{\partial g[w(x,\alpha),\alpha,\beta]}{\partial w(x,\alpha)}.
    \]

    For $\alpha \in (1,2),\beta \in (-\infty,0) \cup (0,1)$ and $x\in (0,\frac{\pi}{16})$, it follows from Lemma \ref{lem3} and Lemma \ref{lem7} that
    \[
    \frac{\partial^{2}g[w(x,\alpha),\alpha,\beta]}{\partial[w(x,\alpha)]^{2}}>0, \left[\frac{\partial w(x,\alpha)}{\partial x}\right]^{2}>0,\frac{\partial^{2} w(x,\alpha)}{\partial x^{2}}<0,\frac{\partial g[w(x,\alpha),\alpha,\beta]}{\partial
    w(x,\alpha)}<0.
    \]
    Then we have
    \[
    \frac{\partial^{2}N(x,\alpha,\beta)}{\partial x^{2}}>0,
    \]
    and therefore
    \[
    \frac{\partial K(x,\alpha,\beta)}{\partial x}=\frac{\partial N(x+\frac{\pi}{32},\alpha,\beta)}{\partial x}-\frac{\partial N(x,\alpha,\beta)}{\partial x} >0,
    \]
    which implies that $K(x,\alpha,\beta)$ is strictly monotonically increasing with respect to $x \in (0,\frac{\pi}{16}]$. Hence we have
    \[
    K\left(\frac{\pi}{64},\alpha,\beta\right)>K(0,\alpha,\beta)=0,
    \]
    which completes the proof. $\hfill\qedsymbol$

\end{document}